\documentclass[showpacs,preprint,prb,aps]{revtex4}
\usepackage{amssymb}
\usepackage{epsfig}
\begin{document}
\title{Quantum analysis of a linear DC SQUID mechanical displacement detector }

\author{M. P. Blencowe}\affiliation{Department of Physics and Astronomy, Dartmouth College, Hanover, New Hampshire
03755 }
\author{E. Buks} \affiliation{Department of Electrical Engineering, Technion, Haifa 32000 Israel}

\date{\today}

\begin{abstract}
We provide a quantum analysis of a DC SQUID mechanical displacement detector within the sub-critical Josephson current  regime. A segment of the SQUID loop forms the mechanical resonator and motion of the latter is transduced inductively through changes in the flux threading the loop.  Expressions are derived for the detector signal response and noise, which are  used to evaluate the position and force detection sensitivity. We also investigate cooling of the mechanical resonator due to detector back reaction. 
\end{abstract}


\maketitle

\section{Introduction}
In a series of recent experiments~\cite{knobel,lahaye,naik} and related theoretical work,\cite{blencoweapl00,zhangjap02,blencowepr04,mozyrskyprl04,clerknjp05,blencowenjp05} it was demonstrated that a displacement detector based on either a normal or superconducting single electronic transistor (SSET) can resolve the motion of a micron-scale mechanical resonator close to the quantum limit as set by Heisenberg's Uncertainty Principle.\cite{caves,braginsky92,clerkprb04} The displacement transduction was achieved by capacitively coupling the gated mechanical resonator to the SSET metallic island. When the resonator is voltage biased, motion of the latter changes the island charging energy and hence the Cooper pair tunnel rates. The resulting modulation in the source-drain tunnel current through the SSET is then read out as a signature of the mechanical motion. 

Given the success of this \emph{capacitive}-based transduction method in approaching the quantum limit, it is natural to consider complementary, \emph{inductive}-based transduction methods in which, for example, a  superconducting quantum interference device (SQUID) is similarly used as an intermediate quantum-limited stage between the micron-scale mechanical resonator and secondary amplification stages.\cite{sembanjp07,buksprb06,mizelprl06,buksprpt06} Unavoidable, fundamental noise sources and how they affect  the SSET and SQUID devices are not necessarily the same. Furthermore, achievable coupling strengths between each type of device and a micron-scale mechanical resonator may be different.  Therefore, it would be interesting to address the merits of the SQUID in comparison with the established SSET for approaching the quantum limit of displacement detection.  

In the present paper, we analyze a DC SQUID-based displacement detector. The SQUID is integrated with a mechanical resonator in the form of a doubly-clamped beam,  shown schematically in Fig.~\ref{schemefig}. Motion of the beam changes the magnetic flux $\Phi$ threading the SQUID loop, hence modulating the current circulating the loop. We shall address the operation of the SQUID displacement detector in the regime for which the loop current is smaller than the Josephson junction critical current $I_c$ and at temperatures well below the superconducting critical temperature. We thus assume that  resistive (normal) current flow through the junctions and accompanying current noise can be neglected. (See for example Ref.~\onlinecite{kochprl80} for a quantum noise analysis of resistively shunted Josephson junctions and Ref.~\onlinecite{kochapl81} for a related analysis of the DC SQUID.) Such an assumption cannot be made with the usual mode of operation for the SSET devices, where the tunnel current unavoidably involves the quasiparticle decay of Cooper pairs, resulting in shot noise. 

As noise source, we will consider the quantum electromagnetic fluctuations within the pump/probe feedline and also transmission line resonator that is connected to the SQUID. This noise is a consequence of the necessary dissipative coupling to the outside world and affects the mechanical signal output in two ways. First, the noise is added directly to the output in the probe line and, second, the noise acts back on the mechanical  resonator via the SQUID, affecting the resonator's motion. 

With the Josephson junction plasma frequencies assumed to be much larger than the other resonant modes of relevance for the device, the SQUID can be modeled to a good approximation as an effective inductance that depends on the external current $I$ entering and exiting the loop, as well as on the applied flux. In this first of two papers, we shall make the further approximation of neglecting the $I$-dependence of the SQUID effective inductance, which requires the condition $I\ll I_c$. In the sequel,\cite{nation??}  we  will  relax this condition somewhat  by including the next to leading $O(I^2)$ term in the inductance and address the consequences of this non-linear correction for quantum-limited displacement detection.

Modeling the SQUID approximately as a passive inductance element, the transmission line resonator-mechanical resonator effective Hamiltonian is given by Eq.~(\ref{hamiltonian3eq}). This Hamiltonian describes many other detector-oscillator systems that are modeled as two coupled harmonic oscillators, including the examples of an $LC$ resonator capacitively coupled to a mechanical resonator~\cite{wineland06,truitt07} and an optical cavity coupled to a mechanically compliant mirror via radiation pressure;\cite{pacepra93,mancinipra94,milburnpra94,jacobspra99,marquardtprpt07} the various systems  are distinguished only by the dependences of the coupling strengths on the parameters particular to each system. Thus, many of the results of this paper are of more general relevance.    

The central results of the paper are Eqs.~(\ref{completesignaleq}) and (\ref{completenoiseeq}), giving the detector response to a mechanical resonator undergoing quantum Brownian motion and also subject to a classical driving force. In the derivation of these expressions, we do not approximate the response as a perturbation series in the coupling between the SQUID and mechanical resonator as is conventionally done, but rather find it more natural to base our approximations instead on assumed weak coupling between the mechanical resonator and its external heat bath and weak classical driving force. Thus, in the context of the linear response paradigm, our detector should properly be viewed as including the mechanical resonator degrees of freedom as well, with the weak perturbative signal instead consisting of the heat bath force noise and classical drive force acting on the mechanical resonator. Since the quality factors of actual, micron-scale mechanical resonators can be very large at sub-Kelvin temperatures (E.g., $Q\sim 10^5$ in the experiments of Refs.~\onlinecite{lahaye,naik}), quantum electromagnetic noise in the transmission line part of the detector can have strong back reaction effects  on the motion of the mechanical resonator, even when the coupling between the resonator and the SQUID is very weak. One consequence  that we shall consider is cooling of the mechanical resonator fundamental mode, which requires strong back reaction damping combined with low noise. Nevertheless, as we will also show, one can still analyze the quantum-limited detector linear response to the mechanical resonator's position signal using general expressions~(\ref{completesignaleq}) and (\ref{completenoiseeq}), under the appropriate conditions of small pump drive and weak coupling  between the SQUID and mechanical resonator such that back reaction effects are small.     

The outline of the paper is as follows. In Sec.~\ref{motionsec}, we write down the SQUID-mechanical resonator equations of motion corresponding to the circuit scheme shown in Fig.~\ref{schemefig} and then derive the Heisenberg equations for the various mode raising and lowering operators, subject to the above-mentioned approximations. In Sec.~\ref{solvesec}, we solve the equations within the linear response approximation to derive the detector signal response and noise. In Sec.~\ref{results}, we analyze both the position and force detection sensitivity, and address also back reaction cooling of the mechanical resonator. Sec.~\ref{conclusion} provides concluding remarks.   

\section{Equations of Motion}
\label{motionsec}

\subsection{Transmission line-SQUID-mechanical oscillator Hamiltonian}
\begin{figure}[htbp]
\centering
\epsfig{file=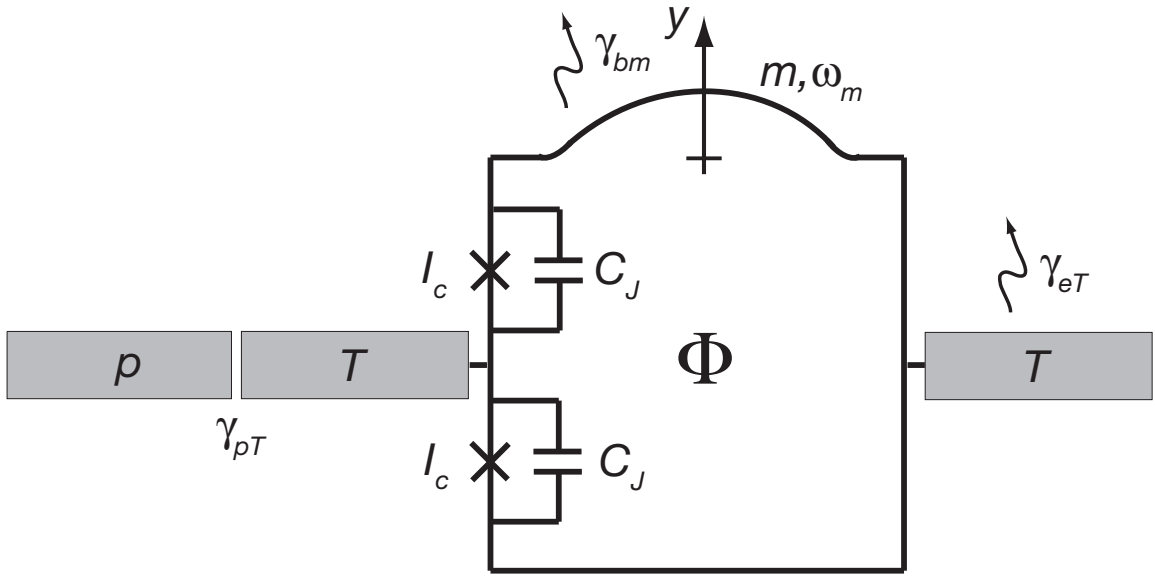,width=6.in} 
\caption{Scheme for the displacement detector showing the pump/probe line `$p$', transmission line resonator `$T$', and DC SQUID with mechanically compliant loop segment having effective mass $m$ and fundamental frequency $\omega_m$. Note that the scale of the DC SQUID is exaggerated relative to that of the stripline for clarity.}  
\label{schemefig}
\end{figure}

Fig.~\ref{schemefig} shows the displacement detector scheme. The device consists of a stripline resonator (transmission line $T$) made of two sections, each of length $l/2$, connected via a DC SQUID (see  Refs.~\onlinecite{blaispra04,wallraffnature04,lupascuprl06,johanssonjpcm06} for related, qubit detection schemes). The transmission line  inductance and capacitance per unit length are $L_T$ and $C_T$ respectively.  The Josephson junctions in each arm of the SQUID are assumed to have identical critical currents $I_c$ and capacitances $C_J$. A length $l_{\mathrm{osc}}$ segment  of the SQUID loop is free to vibrate as a doubly-clamped bar resonator and the fundamental flexural mode of interest (in the plane of the loop) is treated as a harmonic oscillator with mass $m$, frequency $\omega_m$ and displacement coordinate $y$. The total external magnetic flux applied perpendicular to the SQUID loop is given by $\Phi_{\mathrm{ext}}+\lambda B_{\mathrm{ext}}l_{\mathrm{osc}}y$, where $\Phi_{\mathrm{ext}}$ is the flux corresponding to the case $y=0$, $B_{\mathrm{ext}}$ is the normal component of the magnetic field at the location of the vibrating loop segment (oscillator), and the dimensionless parameter   $\lambda<1$ is a geometrical correction factor accounting for the non-uniform displacement of the doubly-clamped resonator in the fundamental flexural mode. 

The transmission line is weakly coupled to a pump/probe feedline ($p$),  with inductance and capacitance per unit length $L_p$ and $C_p$ respectively,  employed for delivering the input and output RF signals; the coupling can be characterized by a transmission line mode amplitude damping rate $\gamma_{pT}$    (see section~\ref{opensec} below). Other possible damping mechanisms in the transmission line may be taken into account by adding a fictitious semi-infinite stripline environment ($e$), weakly coupled to the transmission line characterized by mode amplitude damping rate $\gamma_{eT}$.\cite{yurkepra84} While $\gamma_{eT}$ can be made much smaller than $\gamma_{pT}$ with suitable transmission line resonator design, we shall nevertheless include both sources of damping in our analysis so as to eventually be able to gauge their relative effects on the detector displacement sensitivity [see Eq.~(\ref{noiseoptimumeq})]. The SQUID, on the other hand, is assumed to be dissipationless. The mechanical oscillator is also assumed to be coupled to an external heat bath ($b$), characterized by mode amplitude damping rate $\gamma_{bm}$.       

A convenient choice of dynamical coordinates for the SQUID are $\gamma_{\pm}=\left({\phi_1\pm\phi_2}\right)/2$, where $\phi_1$ and $\phi_2$ are the gauge invariant phases across each of the two Josephson junctions.\cite{orlando91}  For the transmission line, we similarly use its phase field coordinate $\phi(x,t)$,\cite{devoret97,johanssonjpcm06} where $x$ describes the longitudinal location along the transmission line:  $-l/2<x<l/2$, with the SQUID located at $x=0$.  In terms of $\phi$, the transmission line current and voltage are 
\begin{equation}
I_T(x,t)=-\frac{\Phi_0}{2\pi L_T}\frac{\partial\phi(x,t)}{\partial x}
\label{transcurrenteq}
\end{equation}
and
\begin{equation}
V_T(x,t)=\frac{\Phi_0}{2\pi}\frac{\partial\phi(x,t)}{\partial t},
\label{transvoltageeq}
\end{equation}
where $\Phi_0=h/(2e)$ is the flux quantum. Neglecting for now the couplings to the feedline, stripline and mechanical oscillator environments, the equations of motion for the closed system comprising the superconducting transmission line-SQUID-mechanical oscillator are as follows (see, e.g., Ref.~\onlinecite{buksprb06} for a derivation of related equations of motion for a mechanical rf-SQUID): 
\begin{equation}
\frac{\partial^2\phi}{\partial t^2}=(L_TC_T)^{-1}\frac{\partial^2\phi}{\partial x^2},
\label{waveeq}
\end{equation}
\begin{equation}
\omega^{-2}_J\ddot{\gamma}_{-}+\cos(\gamma_{+})\sin(\gamma_{-})+2\beta_L^{-1}\left[\gamma_{-}-\pi\left(n+\frac{(\Phi_{\mathrm{ext}}+\lambda B_{\mathrm{ext}}l_{\mathrm{osc}}y)}{\Phi_0}\right)\right]=0,
\label{phi-eq}
\end{equation}
\begin{equation}
\omega^{-2}_J\ddot{\gamma}_{+}+\sin(\gamma_{+})\cos(\gamma_{-})-\frac{I_T}{2 I_c}=0,
\label{phi+eq}
\end{equation}
and
\begin{equation}
m\ddot{y}+m\omega_m^2 y-\frac{\Phi_0}{\pi L}\lambda B_{\mathrm{ext}}l_{\mathrm{osc}}\gamma_{-}=0,
\label{yeq}
\end{equation}
where  $\omega_J =\sqrt{2\pi I_c/(C_J\Phi_0)}$ is the plasma frequency of the SQUID Josephson junctions, the dimensionless parameter $\beta_L=2\pi LI_c/\Phi_0$,  $L$ is the self inductance of the SQUID,  $n$ is an integer arising from the single-valuedness condition for the phase $2 \gamma_{-}$ around the loop, and $I_T$ is shorthand for $I_T(x=0,t)$. Eq.~(\ref{waveeq}) is simply the wave equation for the phase field coordinate $\phi(x,t)$ of the transmission line. Eq.~(\ref{phi-eq}) describes the current circulating the loop, which depends on the external flux threading the loop. Eq.~(\ref{phi+eq}) describes the average current threading the loop, which from current conservation is equal to one-half the transmission line current at $x=0$. With the circulating SQUID current given  by ${\Phi_0\gamma_-}/(\pi L)$ (up to a $\Phi_{\mathrm{ext}}$ dependent term), we recognize in Eq.~(\ref{yeq})  the Lorentz force acting on the mechanical oscillator.

In addition to the equations of motion, we have the following current and voltage boundary conditions:
\begin{equation}
I_T(x=\pm l/2,t)=0
\label{currentbceq}
\end{equation}
and
\begin{equation}
\frac{\partial\left({L_{\mathrm{eff}}\left[\Phi_{\mathrm{ext}}(y),I_T\right]}I_T\right)}{\partial t}=V_T(0^-,t)-V_T(0^+,t),
\label{voltagebceq}
\end{equation}
where  the external flux and current-dependent, effective inductance $L_{\mathrm{eff}}\left[\Phi_{\mathrm{ext}}(y),I_T\right]$ of the SQUID as `seen' by the transmission line is
\begin{equation}
L_{\mathrm{eff}}\left[\Phi_{\mathrm{ext}}(y),I_T\right]=\frac{\Phi_0 \gamma_+}{2\pi I_T} +\frac{L}{4},
\label{effectiveinductanceeq}
\end{equation}
with $\Phi_{\mathrm{ext}}(y)=\Phi_{\mathrm{ext}}+\lambda B_{\mathrm{ext}} l_{\mathrm{osc}} y$. Note that we have set $n=0$, since observable quantities do not depend on $n$. 

We now make the following assumptions and consequent approximations: (a) $\omega_J\gg\omega_T\gg\omega_m$ (where $\omega_T$ is the relevant resonant mode of the transmission line); neglect the SQUID inertia terms $\omega_J^{-2}\ddot{\gamma}_{\pm}$. (b) $\beta_L\ll 1$; solve for $\gamma_{\pm}$ as series expansions to first order in $\beta_L$. (c) $\left|B_{\mathrm{ext}} l_{\mathrm{osc}} y\right|/\Phi_0\ll1$; series expand the equations of motion to first order in $y(t)$. (d) $\left|I_T/I_c\right|=\left|\frac{\Phi_0}{2\pi L_T I_c}\frac{\partial\phi (0,t)}{\partial x}\right|\ll 1$;  series expand the equations of motion to second order in $I_T$.  

With $\omega_J$'s typically in the tens of GHz, assumption~(a) is reasonable. From Eq.~(\ref{phi-eq}), we see that a small $\beta_L$ value prevents the $\gamma_-$ coordinate from getting trapped in its various potential minima, causing unwanted hysteresis. With the $\gamma_+$ expansion in $I_T$ consisting of only odd powers, approximations (a) and (d)  amount  to describing the SQUID simply  as a current independent, $\Phi_{\mathrm{ext}}$-tunable passive  inductance element $L_{\mathrm{eff}}\left[\Phi_{\mathrm{ext}}(y)\right]$  that also depends on the mechanical oscillator position coordinate $y$.  Including the next-to-leading, $I_T^3$ term in the $\gamma_+$ expansion gives an $I^2_T$-dependent,  nonlinear correction to the SQUID effective inductance. The consequences of including this nonlinear correction term for the quantum-limited displacement detection sensitivity will be considered in a forthcoming paper.\cite{nation??} Solving for $\gamma_+$  to order $I_T$ and substituting in Eq.~(\ref{effectiveinductanceeq}), we obtain: 
\begin{equation}
L_{\mathrm{eff}}\left[\Phi_{\mathrm{ext}}(y)\right]\approx\frac{\Phi_0}{4\pi I_c}\sec\left(\frac{\pi\Phi_{\mathrm{ext}}(y)}{\Phi_0}\right),
\label{effectiveinductanceeq2}
\end{equation}
where the self inductance $L$ contribution has been neglected since it is of order $\beta_L\ll1$. 
Solving for $\gamma_-$ to order $I_T^2$ and substituting into Eq.~(\ref{yeq}), we obtain for the mechanical oscillator equation of motion:
\begin{equation}
m\ddot{y}+m\omega_m^2 y-\frac{\pi\lambda B_{\mathrm{ext}}l_{\mathrm{osc}} I_T^2}{8I_c}\tan\left(\pi\Phi_{\mathrm{ext}}/\Phi_0\right)\sec\left(\pi\Phi_{\mathrm{ext}}/\Phi_0\right)=0,
\label{oscillatoreq}
\end{equation}
where from (c), we have set $y=0$ in the solution for $\gamma_-$ and have dropped an overall constant term. 
Since  the  $\gamma_-$ expansion in $I_T$ consists only of even powers, we must go  to second order in $I_T$ so as to have a non-trivial transmission line-oscillator effective coupling. Thus, the SQUID phase coordinates $\gamma_{\pm}$ have been completely eliminated from the equations of motion, a consequence of approximation (a); the SQUID mediates the interaction between the mechanical oscillator coordinate $y$ and transmission line coordinate $\phi$ without retardation effects.  

From Eq.~(\ref{oscillatoreq}), it might appear that the force on the mechanical oscillator due to the transmission line can be made arbitrarily large by tuning $\Phi_{\mathrm{ext}}$ close to $\Phi_0/2$. Note, however, that the proper conditions for the validity of the $I_T$ and $\beta_L$ expansions are:
\begin{equation}
\left|\frac{I_T}{I_c}\sec\left(\pi\Phi_{\mathrm{ext}}/\Phi_0\right)\right|\ll 1
\label{currentcondition}
\end{equation}
and 
\begin{equation}
\left|\beta_L\sec\left(\pi\Phi_{\mathrm{ext}}/\Phi_0\right)\right|\ll 1.
\label{betacondition}
\end{equation}  

We now restrict ourselves to a single transmission line mode and derive approximate equations of motion for the mode amplitude. Suppose that the mechanical oscillator position coordinate is held fixed at $y=0$.  The following phase field satisfies the current boundary conditions~(\ref{currentbceq}): 
\begin{equation}\phi(x,t)=\left\{
\begin{array}{cc}
  -\phi(t)\cos\left[k_0\left(x+l/2\right)\right]; &x<0  \\
   +\phi(t)\cos\left[k_0\left(x-l/2\right)\right]; &x>0
\end{array}\right.,
\label{currentbc2eq}
\end{equation}
with the wavenumber $k_0$ determined by the voltage boundary condition~(\ref{voltagebceq}):
 \begin{equation}
\frac{k_0 l}{2}\tan\left(\frac{k _0 l}{2}\right)=-\frac{ L_T  l}{L_{\mathrm{eff}}\left(\Phi_{\mathrm{ext}}\right)}.
\label{wavenoceq}
\end{equation}
The wave equation~(\ref{waveeq}) gives for the transmission mode frequency: $\omega_T=k_0/\sqrt{L_T C_T}$.
Substituting the phase field~(\ref{currentbc2eq}) into the $I_T$ part of the oscillator equation of motion~(\ref{oscillatoreq}) furthermore gives the transmission line force acting on the oscillator with fixed coordinate $y=0$. Now release the mechanical oscillator coordinate and suppose that for small [condition~(c)] , slow [condition~(a)] displacements, the force is the same to a good approximation.Then the oscillator equation of motion becomes
\begin{eqnarray}
&&m\ddot{y}(t)+m\omega_m^2 y(t)+\frac{1}{4}C_Tl \left(\frac{\Phi_0}{2\pi}\right)^2\sin^2\left(k_0 l/2\right)\cr
&&\times\left[-\frac{\lambda B_{\mathrm{ext}} l_{\mathrm{osc}}}{\left(\Phi_0/2\pi\right)}\cdot\frac{\Phi_0}{4\pi L_T l I_c}\tan\left(\pi\Phi_{\mathrm{ext}}/\Phi_0\right)\sec\left(\pi\Phi_{\mathrm{ext}}/\Phi_0\right)\right]\omega_T^2\phi^2(t)=0,
\label{oscillator2eq}
\end{eqnarray}
From Eq.~(\ref{oscillator2eq}), we can determine the mechanical sector of the Lagrangian, along with the interaction potential involving $y$ and the mode amplitude $\phi$. The remaining transmission line sector follows from the wave equation~(\ref{waveeq}) and we thus have for the total Lagrangian: 
\begin{eqnarray}
&&L\left(\phi,y,\dot{\phi},\dot{y}\right)=\frac{1}{2}m\dot{y}^2-\frac{1}{2}m\omega_m^2 y^2+\frac{1}{2}C_Tl \left(\frac{\Phi_0}{2\pi}\right)^2\sin^2\left(k_0 l/2\right)\cr
&&\times\left\{\frac{1}{2}\dot{\phi}^2-\frac{1}{2}\left[1-\frac{\lambda B_{\mathrm{ext}} l_{\mathrm{osc}}y}{\left(\Phi_0/2\pi\right)}\cdot\frac{\Phi_0}{4\pi L_T l I_c}\tan\left(\pi\Phi_{\mathrm{ext}}/\Phi_0\right)\sec\left(\pi\Phi_{\mathrm{ext}}/\Phi_0\right)\right]\omega_T^2\phi^2\right\}.
\label{langrangianeq}
\end{eqnarray}
From Eq.~(\ref{langrangianeq}), we see that for motion occuring on the much longer timescale $\omega_m^{-1} \gg\omega_T^{-1}$, the  mechanical oscillator has the effect of modulating the frequency of the transmission line mode.  

The associated Hamiltonian is
\begin{eqnarray}
&&H\left(\phi,y,p_{\phi},p_y\right)=
\left[\frac{2}{C_Tl \left(\frac{\Phi_0}{2\pi}\right)^2\sin^2\left(k_0 l/2\right)}\right]\frac{1}{2} p_{\phi}^2
+\frac{1}{2}C_Tl \left(\frac{\Phi_0}{2\pi}\right)^2\sin^2\left(k_0 l/2\right)\cr
&&\times\left[1-\frac{\lambda B_{\mathrm{ext}} l_{\mathrm{osc}}y}{\left(\Phi_0/2\pi\right)}\cdot\frac{\Phi_0}{4\pi L_T l I_c}\tan\left(\pi\Phi_{\mathrm{ext}}/\Phi_0\right)\sec\left(\pi\Phi_{\mathrm{ext}}/\Phi_0\right)\right]\frac{1}{2}\omega_T^2\phi^2\cr
&&+\frac{p_y^2}{2 m} +\frac{1}{2}m\omega_m^2 y^2.
\label{hamiltonianeq}
\end{eqnarray}

Let us now quantize. For the transmission line mode coordinate, the raising(lowering) operator is defined as:
\begin{equation}\hat{a}^{\pm}_T=\frac{1}{\sqrt{2\hbar\omega_T \left[\frac{1}{2}C_T l\left(\Phi_0/2\pi\right)^2\sin^2\left(k_0 l/2\right)\right]}}\left[\frac{1}{2}C_Tl \left(\frac{\Phi_0}{2\pi}\right)^2\sin^2\left(k_0 l/2\right) \omega_T \hat{\phi}\mp i \hat{p}_{\phi}\right]
\label{transoperatoreq}
\end{equation}
and for the mechanical oscillator
\begin{equation}
\hat{a}^{\pm}_m=\frac{1}{\sqrt{2m\omega\hbar}} \left(m\omega \hat{y}\mp  i\hat{p}_y\right).
\label{mechoperator}
\end{equation}
In terms of these operators, the Hamiltionian (\ref{hamiltonianeq}) becomes (for notational convenience we omit from now on the `hats' on the operators and also the `minus' superscript on the lowering operator):
\begin{eqnarray}
H=\hbar \omega_T a^+_T a_T +\hbar \omega_m a^+_m a_m +\frac{1}{2}\hbar\omega_T K_{Tm}\left(a_T +a_T^+\right)^2 \left(a_m +a_m^+\right),
\label{hamiltonian2eq}
\end{eqnarray}
where the dimensionless coupling parameter between the mechanical oscillator and transmission line mode is
\begin{equation}
K_{Tm}=-\frac{\lambda B_{\mathrm{ext}} l_{\mathrm{osc}}\Delta x_{zp}}{\left(\Phi_0/2\pi\right)}\frac{\Phi_0}{4\pi L_T l I_c}\tan\left(\pi\Phi_{\mathrm{ext}}/\Phi_0\right)\sec\left(\pi\Phi_{\mathrm{ext}}/\Phi_0\right),
\label{coupling}
\end{equation}
with $\Delta x_{zp}=\sqrt{\hbar/(2m\omega_m)}$  the zero-point uncertainty of the mechanical oscillator. 
From expression~(\ref{effectiveinductanceeq2}) for the effective inductance, another way to express the coupling parameter is as follows:
\begin{equation}
K_{Tm}=-\frac{\lambda B_{\mathrm{ext}} l_{\mathrm{osc}}\Delta x_{zp}}{\left(\Phi_0/2\pi\right)}  \frac{\Phi_0}{\pi}\frac{d L_{\mathrm{eff}}/d\Phi_{\mathrm{ext}}}{L_T l}.
\label{coupling2}
\end{equation}
From Eq.~(\ref{coupling2}), we see that in order to increase the coupling between the mechanical oscillator and transmission line, the  SQUID effective inductance-to-transmission line inductance ratio must be increased. The advantage of using a SQUID over an ordinary, geometrical mutual inductance between a transmission line and micron-sized mechanical oscillator is that the former can give a much larger effective inductance.  As we shall see in Sec.~\ref{results}, just requiring that the inductances be matched such that $ \frac{\Phi_0}{\pi}\frac{d L_{\mathrm{eff}}/d\Phi_{\mathrm{ext}}}{L_T l}\sim 1$ is sufficient for strong back reaction effects with modest drive powers, even though the other term in $K_{Tm}$ describing the flux induced for a zero-point displacement is typically very small.

Assuming then that  $K_{Tm}\ll1$ and making the rotating wave approximation (RWA) for the `$T$' part of the interaction term in the system Hamiltonian~(\ref{hamiltonian2eq}), i.e., neglecting the terms $(a_T)^2$ and $(a_T^{+})^2$, we have (up to an unimportant additive constant):
\begin{equation}
H=\hbar \omega_T a^+_T a_T +\hbar \omega_m a^+_m a_m +\hbar\omega_T K_{T m}a_T^+a_T\left(a_m +a_m^+\right).
\label{hamiltonian3eq}
\end{equation}
Many other systems are modeled by this form of Hamiltonian, a notable example being the single mode of an optical cavity  interacting via radiation pressure with a mechanically compliant mirror.\cite{pacepra93,mancinipra94,milburnpra94,jacobspra99,marquardtprpt07} Thus, much of the subsequent analysis will be relevant to a broad class of coupled resonator devices--not to just the transmission line-SQUID-mechanical resonator system.

\subsection{Open system Heisenberg equations of motion}
\label{opensec}
So far, we have treated the transmission line  and mechanical resonator as a closed system with SQUID-induced effective coupling . Of course, a real transmission line mode will experience damping and accompanying fluctuations, not least because it must be coupled to the outside world in order for its state to be measured. Furthermore, the mechanical resonator mode will of course be damped even when decoupled from the SQUID. It is straightforward to incorporate the various baths and pump/probe feedline in terms of raising/lowering operators. Assuming weak system-bath couplings, which again justify the RWA, we have for the full Hamiltonian:
\begin{eqnarray}
H&=&\hbar \omega_T a^+_T a_T +\hbar \omega_m a^+_m a_m +\hbar\omega_T K_{T m}a_T ^+a_T \left(a_m +a_m^+\right) \cr
&&+\hbar\int d\omega \omega a_p^+(\omega)a_p(\omega)+\hbar\int d\omega \omega a_e^+(\omega)a_e(\omega)+\hbar\int d\omega \omega a_b^+(\omega)a_b(\omega)\cr
&&+\hbar\int d\omega \left[K^*_{p  T}a_p^+ (\omega) a_T+K_{p  T}a_T^+ a_p (\omega)\right]+
\hbar\int d\omega \left[K^*_{e  T}a_e^+ (\omega) a_T+K_{p  T}a_T^+ a_e (\omega)\right]\cr
&&+\hbar\int d\omega \left[K^*_{b m}a_b^+ (\omega) a_m+K_{b m}a_m^+ a_b (\omega)\right]
-\sqrt{\frac{\hbar}{2 m\omega_m}} (a_m +a_m^+) F_{\mathrm{ext}}(t),
\label{hamiltonianopeneq}
\end{eqnarray}
where $a_p$ denotes the pump/probe ($p$) feed line operator, $a_e$ the transmission line bath (`$e$' for `environment') operator, and $a_b$ the  mechanical resonator bath ($b$) operator. These operators satisfy the usual canonical commutation relations:
\begin{equation}
\left[a_i(\omega),a_j^+(\omega')\right]=\delta_{ij}\delta(\omega-\omega').
\label{cr1eq}
\end{equation}
The couplings between these baths and the transmission line and mechanical resonator systems are denoted as $K_{p  T}$, $K_{e  T}$, and $K_{b m}$. Note we have also included for generality a classical driving force $F_{\mathrm{ext}}(t)$ acting on the mechanical resonator. This allows us the opportunity to later on analyze quantum limits on force detection in addition to displacement detection. 

Within the RWA, it is straightforward to solve the Heisenberg equations for the bath operators and substitute these solutions into the Heisenberg equations for the transmission line and mechanical oscillator to give
\begin{eqnarray}
\frac{da_m}{dt}&=&-i\omega_m a_m +\frac{i}{\hbar}\sqrt{\frac{\hbar}{2m\omega_m}} F_{\mathrm{ext}}(t)-i\omega_T K_{T m} a_T^+a_T\cr
&&-\int d\omega \left|K_{T m}\right|^2\int_{t_0}^t dt' e^{-i \omega(t-t')}a_m(t')-i\int d\omega K_{b m} e^{-i\omega (t-t_0)} a_b (\omega,t_0)
\label{heisenmeq}
\end{eqnarray}
and
\begin{eqnarray}
\frac{da_T}{dt}&=&-i\omega_T a_T -i\omega_T K_{T m} a_T\left(a_m+a_m^+\right)\cr
&&-\int d\omega \left|K_{p  T}\right|^2\int_{t_0}^t dt' e^{-i \omega(t -t')}a_T(t')-i\int d\omega K_{p  T} e^{-i\omega (t-t_0)} a_p (\omega,t_0)\cr
&&-\int d\omega \left|K_{e  T}\right|^2\int_{t_0}^t dt' e^{-i \omega(t -t')}a_T(t')-i\int d\omega K_{e  T} e^{-i\omega (t -t_0)} a_e (\omega,t_0).
\label{heisenTeq}
\end{eqnarray}

We now make the so-called `first Markov approximation',\cite{gardinerpra85,gardiner00} in which the frequency dependences of the couplings to the baths are neglected:
\begin{eqnarray}
K_{p  T}(\omega)&=&\sqrt{\frac{\gamma_{p  T}}{\pi}}e^{i\phi_{p  T}}\cr
K_{e  T}(\omega)&=&\sqrt{\frac{\gamma_{e  T}}{\pi}}e^{i\phi_{e  T}}\cr
K_{b m}(\omega)&=&\sqrt{\frac{\gamma_{b m}}{\pi}}e^{i\phi_{b m}},
\end{eqnarray}
where the $\gamma$'s and $\phi$'s are independent of $\omega$ as stated. The Heisenberg equations of motion (\ref{heisenmeq}) and (\ref{heisenTeq}) then simplify to 
\begin{eqnarray}
\frac{da_m}{dt}&=&-i\omega_m a_m +\frac{i}{\hbar}\sqrt{\frac{\hbar}{2m\omega_m}} F_{\mathrm{ext}}(t)-i\omega_T K_{T m}a_T^+a_T\cr
&&-\gamma_{b m} a_m(t)-i\sqrt{2 \gamma_{b m}}e^{i\phi_{b m}}a_b^{\mathrm{in}}(t)
\label{heisenm2eq}
\end{eqnarray}
and
\begin{eqnarray}
\frac{da_T}{dt}&=&-i\omega_T a_T -i\omega_T K_{T m} a_T\left(a_m+a_m^+\right)\cr
&&-\gamma_{p  T}a_T(t)-i\sqrt{2\gamma_{p  T}}e^{i\phi_{p  T}}a_p^{\mathrm{in}} (t)\cr
&&-\gamma_{e  T}a_T(t)-i\sqrt{2\gamma_{e  T}}e^{i\phi_{e  T}}a_e^{\mathrm{in}} (t),
\label{heisenT2eq}
\end{eqnarray}
where the $\gamma_i$'s are the various mode amplitude damping rates (assumed much smaller than their associated mode frequencies) and  the `in' operators~\cite{caves,yurkepra84,gardinerpra85,gardiner00} are defined as
\begin{equation}
a_i^{\mathrm{in}}(t)=\frac{1}{\sqrt{2\pi}}\int d\omega e^{-i\omega (t-t_0)}a_i (\omega,t_0),
\end{equation}
with $t>t_0$. The time $t_0$ can be taken to be an instant in the distant past before the measurement commences and when the initial conditions are specified (see below).
We can similarly define `out' operators:
\begin{equation}
a_i^{\mathrm{out}}(t)=\frac{1}{\sqrt{2\pi}}\int d\omega e^{-i\omega (t-t_1)}a_i (\omega,t_1),
\end{equation}
 with $t_1>t$. The time $t_1$ can be taken to be an instant in the distant future after the measurement has finished. From the Heisenberg equations for the bath  operators and the definitions of the `in' and `out' operators, we obtain the following identities between them:\cite{gardinerpra85,gardiner00}
\begin{eqnarray}
a_p^{\mathrm{out}}(t)-a_p^{\mathrm{in}}(t)&=&-i\sqrt{2\gamma_{pT}}e^{-i\phi_{pT}}a_T(t)\cr
a_b^{\mathrm{out}}(t)-a_b^{\mathrm{in}}(t)&=&-i\sqrt{2\gamma_{bm}}e^{-i\phi_{bm}}a_m(t)\cr
a_e^{\mathrm{out}}(t)-a_e^{\mathrm{in}}(t)&=&-i\sqrt{2\gamma_{eT}}e^{-i\phi_{eT}}a_T(t).
\label{inoutidentityeq}
\end{eqnarray}

In outline, the method of solution runs in principle as follows:\cite{yurkepra84,gardinerpra85,gardiner00,yurkejlt07} (1) specify the `in' operators. (2) Solve for the system operators $a_m(t)$ and $a_T(t)$ in terms of the `in' operators. (3) Use the relevant identity (\ref{inoutidentityeq}) to determine the `out' operator $a_p^{\mathrm{out}}(t)$, which yields the desired probe signal. 
It is more convenient to solve the Heisenberg equations in the frequency domain with the Fourier transformed operators $O(t)=\frac{1}{\sqrt{2\pi}}\int_{-\infty}^{\infty} d\omega e^{-i\omega t} O(\omega)$. The equations for the system operators then become
\begin{eqnarray}
a_m(\omega)&=&\frac{1}{\omega-\omega_m +i\gamma_{b m}}\left\{\sqrt{2\gamma_{b m}} e^{i\phi_{b m}} a_b^{\mathrm{in}} (\omega) -\frac{1}{\sqrt{2 m\hbar\omega_m}} F_{\mathrm{ext}}(\omega)\right.\cr
&&\left.+\frac{\omega_T K_{T m}}{2\sqrt{2\pi}}\int_{-\infty}^{\infty} d \omega' \left[a_T (\omega') a_T^+ (\omega'-\omega)+a_T^+ (\omega') a_T (\omega+\omega')\right]\right\}
\label{heisenm3eq}
\end{eqnarray}
and
\begin{eqnarray}
a_T(\omega)&=&\frac{1}{\omega-\omega_T +i(\gamma_{p T}+\gamma_{e T})}\left\{\sqrt{2\gamma_{p T}} e^{i\phi_{p T}} a_p^{\mathrm{in}} (\omega)+\sqrt{2\gamma_{e T}} e^{i\phi_{e T}} a_e^{\mathrm{in}} (\omega)\right.\cr
&&\left.+\frac{\omega_T K_{T m}}{\sqrt{2\pi}}\int_{-\infty}^{\infty} d \omega' a_T (\omega') \left[a_m (\omega-\omega') + a_m^+ (\omega'-\omega)\right]\right\},
\label{heisenT3eq}
\end{eqnarray}
while the relevant `in/out' operator identity becomes
\begin{equation}
a_p^{\mathrm{out}}(\omega)=-i\sqrt{2\gamma_{p T}} e^{-i\phi_{p T}} a_T (\omega) +a_p^{\mathrm{in}} (\omega).
\label{inoutidentity2eq}
\end{equation}

\subsection{Observables and `in' states}
Before proceeding with the solution to Eqs.~(\ref{heisenm3eq}) and (\ref{heisenT3eq}), let us first devote some time to deriving expressions for observables that we actually measure  in terms of $a_p^{\mathrm{out}}(\omega)$. Model the pump/probe feedline as a semi-infinite transmission line $-\infty<x<0$. Solving the wave equation for the decoupled transmission line and then using the expressions (\ref{transcurrenteq}),   (\ref{transvoltageeq}) relating the current/voltage to the phase coordinate, we obtain
\begin{equation}
I^{\mathrm{out}}(x,t)=-\int_{-\infty}^{\infty} d\omega \sqrt{\frac{\hbar\omega}{\pi Z_p}}\sin \left(\omega x/v_p\right)\left[e^{-i\omega t} a_p^{\mathrm{out}}(\omega)+e^{i\omega t} a_p^{{\mathrm{out}} +} (\omega)\right]
\label{outcurrenteq}
\end{equation}
and
\begin{equation}
V^{\mathrm{out}}(x,t)=i\int_{-\infty}^{\infty} d\omega \sqrt{\frac{Z_p\hbar\omega}{\pi}}\cos \left(\omega x/v_p\right)\left[e^{-i\omega t} a_p^{\mathrm{out}}(\omega)-e^{i\omega t} a_p^{{\mathrm{out}}+} (\omega)\right],
\label{outvoltageeq}
\end{equation}
where the sinusoidal $x$ dependence in the current expression follows from the vanishing of the current boundary condition at $x=0$, the feedline impedance is $Z_p=\sqrt{L_p/C_p}$ and the wave propagation velocity is $v_p=1/\sqrt{L_p C_p}$. Suppose the current/volt meter is at $x\rightarrow -\infty$,  so that the actual observables correspond to measuring the left-propagating component of the current/voltage. Then  decomposing the $x$-dependent trig terms into their real and imaginary parts, we can identify the left propagating current/voltage operators as
\begin{eqnarray}
I^{\mathrm{out}}(x,t)&=&-i\sqrt{\frac{\hbar}{4\pi Z_p}}\int_0^{\infty} d\omega\sqrt{\omega}\left[e^{-i\omega(x/v_p+t)} 
\left(a_p^{\mathrm{out}} (\omega)-a_p^{{\mathrm{out}}+}(-\omega)\right)\right. \cr
&&\left.+e^{i\omega (x/v_p+t)}\left(a_p^{\mathrm{out}} (-\omega)-a_p^{{\mathrm{out}}+}(\omega)\right)\right]
\label{outcurrent2eq}
\end{eqnarray}
and
\begin{eqnarray}
V^{\mathrm{out}}(x,t)&=&i\sqrt{\frac{Z_p\hbar}{4\pi}}\int_0^{\infty} d\omega\sqrt{\omega}\left[e^{-i\omega(x/v_p+t)} 
\left(a_p^{\mathrm{out}} (\omega)-a_p^{{\mathrm{out}} +}(-\omega)\right)\right. \cr
&&\left.+e^{i\omega(x/v_p+t)}\left(a_p^{\mathrm{out}} (-\omega)-a_p^{{\mathrm{out}} +}(\omega)\right)\right].
\label{outvoltage2eq}
\end{eqnarray}
The output signal of interest due to the mechanical oscillator signal input will lie within some bandwidth $\delta\omega$ centered at $\omega_s$, the `signal' frequency, and so we define the filtered output current $I^{\mathrm{out}}\left(x,t|\omega_s,\delta\omega\right)$ and voltage $V^{\mathrm{out}}\left(x,t|\omega_s,\delta\omega\right)$ to be the same as the above, left-moving operators, but with the integration range instead restricted to the interval $\left[\omega_s-\delta\omega/2,\omega_s+\delta\omega/2\right]$.

Since the motion of the mechanical resonator modulates the transmission line frequency, one way to transduce displacements is to measure the relative phase shift between the `in' pump current and `out' probe current using the homodyne detection procedure.\cite{gardiner00} Another common way is  to measure the `out' power relative to the `in' power, or equivalently the mean-squared current/voltage (all three quantities differ by trivial factors of $Z_p$). We will discuss the latter method of transduction; the former, homodyne method can be straightforwardly addressed using similar techniques to those presented here. Thus, we consider the following expectation value:
\begin{equation}
\left\langle\left[\delta I^{\mathrm{out}}\left(x,t|\omega_s,\delta\omega\right)\right]^2\right\rangle=\left\langle\left[I^{\mathrm{out}}\left(x,t|\omega_s,\delta\omega\right)\right]^2\right\rangle-\left\langle I^{\mathrm{out}}\left(x,t|\omega_s,\delta\omega\right)\right\rangle^2,
\end{equation}
where the angle brackets denote an ensemble average with respect to the `in' states of the various baths and feedline (see below). If the mechanical oscillator is being driven by a classical external force whose fluctuations are invariant under time translations, i.e., $\left\langle F_{\mathrm{ext}}(t) F_{\mathrm{ext}}(t')\right\rangle= C(t-t')$, then the above, mean-squared current will be time-independent. Alternatively, if $F_{\mathrm{ext}}(t)$ is, e.g., some deterministic, AC drive, then  we must also time-average so as to get a time-independent measure of the detector response:
\begin{equation}
 \overline{ \left\langle\left[\delta I^{\mathrm{out}}\left(x,t|\omega_s,\delta\omega\right)\right]^2\right\rangle}=\frac{1}{T_M}\int_{-T_M /2}^{T_M /2} dt \left\langle\left[ I^{\mathrm{out}}\left(x,t|\omega_s,\delta\omega\right)\right]^2\right\rangle,
\end{equation}
where $T_M$ is duration of the measurement, assumed much larger than all other timescales associated with the detector dynamics. We have also assumed that the time-averaged current vanishes in the signal bandwidth of interest: $\overline{\left\langle I^{\mathrm{out}}\left(\omega_s,\delta\omega\right)\right\rangle}=0$. Substituting in the expression (\ref{outcurrent2eq})  for $I^{\mathrm{out}}\left(x,t|\omega_s,\delta\omega\right)$ in terms of the $a_p^{\mathrm{out}}$ operators, we obtain after some algebra:
\begin{eqnarray}
\overline{ \left\langle\left[\delta I^{\mathrm{out}}\left(\omega_s,\delta\omega\right)\right]^2\right\rangle}&=&\frac{1}{Z_p}\int_{\omega_s -\delta\omega/2}^{\omega_s+\delta\omega/2}\frac{d\omega_1 d\omega_2}{2\pi} \hbar\omega_1\left(\frac{2}{\left(\omega_1-\omega_2\right) T_M}\sin\left[(\omega_1-\omega_2)T_M/2\right]\right)\cr 
&&\times\frac{1}{2}\left\langle a_p^{\mathrm{out}} (\omega_1) a_p^{{\mathrm{out}} +} (\omega_2) +a_p^{{\mathrm{out}}+}(\omega_2) a_p^{\mathrm{out}} (\omega_1)\right\rangle.
\label{outcurrentsquaredeq}
\end{eqnarray}

As `in' states, we suppose $k_B T\ll \hbar\omega_T$, such that the relevant transmission line `in' bath modes ($\omega_e\sim\omega_T$) are assumed to be approximately in the vacuum state. On the other hand,  with the mechanical mode typically at a much lower frequency $\omega_m\ll\omega_T$, we assume that its relevant `in' bath modes ($\omega_b\sim\omega_m$)  are in the proper, non-zero temperature thermal state. For the  pump/probe feedline, we consider the following coherent state:\cite{johanssonjpcm06}
\begin{equation}
\left|\{\alpha(\omega)\}\right\rangle_p=\exp\left[\int d\omega\alpha(\omega) \left(a_p^{{\mathrm{in}}+}(\omega)-a_p^{\mathrm{in}}(\omega)\right)\right]\left|0\right\rangle_p,
\label{coherentsteq}
\end{equation}
where $\left|0\right\rangle_p$ is the vacuum state and 
\begin{equation}
\alpha(\omega)=-I_0 \sqrt{\frac{Z_p T_M^2}{2\hbar}} \frac{e^{-(\omega-\omega_p)^2 T_M^2 /2}}{\sqrt{\omega}},
\end{equation}
normalized such that the amplitude of the expectation value of $I^{\mathrm{in}}$ [the right propagating version of (\ref{outcurrent2eq}) with $a_p^{\mathrm{out}}$ replaced by $a_p^{\mathrm{in}}$]  with respect to this state is just $I_0$. Again, we suppose $k_B T\ll \hbar\omega_p$,  so that  thermal fluctuations of the feedline are neglected. The frequency width of this pump drive is assumed to be  the inverse lifetime of the measurement.  Below we shall see that the output mechanical signal will appear as two `satellite' peaks on either side of the central peak at $\omega_p$ due to the pump signal, i.e, the mechanical signal can be extracted by centering the filter at either of $\omega_s=\omega_p\pm\omega_m$ (up to a renormalization of the mechanical oscillator frequency), corresponding to the anti-Stokes and Stokes bands. 

Note that we do not have to specify the initial $t_0$ states of the mechanical resonator and transmission line systems;  $a_T(t_0)$ and $a_m(t_0)$-dependent initial transients have been dropped in the above equations for $a_T(\omega)$ and $a_m(\omega)$, since they give a negligible contribution to the long-time, steady-state behavior of interest.

\section{Solving the equations of motion}
\label{solvesec}

\subsection{Linear response approximation}
\label{linresponse}
We are now ready to solve for $\overline{\left\langle\left[\delta I^{\mathrm{out}}\right]^2\right\rangle}$. Introduce the following shorthand notation:
\begin{eqnarray}
S_T(\omega)&=&\sqrt{2 \gamma_{p T}}e^{i\phi_{p T}} a_p^{\mathrm{in}}(\omega)+\sqrt{2 \gamma_{eT}}e^{i\phi_{e T}} a_e^{\mathrm{in}}(\omega)\cr
S_m(\omega)&=&\sqrt{2 \gamma_{b m}}e^{i\phi_{b m}} a_b^{\mathrm{in}}(\omega)-\frac{1}{\sqrt{2m\hbar\omega_m}}F_{\mathrm{ext}}(\omega)\cr
{\cal K}&=&\frac{\omega_T K_{T m}}{\sqrt{2\pi}},
\label{shorthandeq}
\end{eqnarray}
and $\gamma_T=\gamma_{p T}+\gamma_{e T}$, the net transmission line mode amplitude dissipation rate due to loss via the probe line and the transmission line bath.
Substituting Eq.~(\ref{heisenm3eq}) for $a_m(\omega)$ into Eq.~(\ref{heisenT3eq})  for $a_T(\omega)$ yields the following, single equation in terms of $a_T(\omega)$ only:
\begin{eqnarray}
a_T (\omega)&=&\int_{-\infty}^{\infty}d\omega'a_T(\omega-\omega')A(\omega,\omega')+\int_{-\infty}^{\infty}d\omega'B(\omega,\omega')a_T(\omega-\omega')\cr
&&\times\int_{-\infty}^{\infty}d\omega'' \left[a_T (\omega'') a_T^+ (\omega'' -\omega')+a_T^+ (\omega'') a_T (\omega''+\omega')\right]+C(\omega),
\label{Teq}
\end{eqnarray}
where, for the convenience of subsequent calculations, we have made this equation as concise as possible with the following definitions:
 \begin{eqnarray}
 A(\omega,\omega')&=&\frac{{\cal K}}{\omega-\omega_T+i\gamma_T}\left[\frac{S_m(\omega')}{\omega'-\omega_m+i\gamma_{b m}}+\frac{S_m^+(-\omega')}{-\omega'-\omega_m-i\gamma_{b m}}\right],\cr
B(\omega,\omega')&=&\frac{{\cal K}^2 /2}{\omega-\omega_T+i\gamma_T}\left[\frac{1}{\omega'-\omega_m+i\gamma_{b m}}+\frac{1}{-\omega'-\omega_m-i\gamma_{b m}}\right],\cr
C(\omega)&=&\frac{S_T(\omega)}{\omega-\omega_T+i\gamma_T}.
\label{shorthand2eq}
\end{eqnarray}

We expand Eq.~(\ref{Teq}) for $a_T(\omega)$ to first order in the mechanical oscillator bath operator $a_b^{\mathrm{in}}(\omega)$ and external driving force $F_{\mathrm{ext}}(\omega)$ [equivalently expand in $A(\omega,\omega')$]:  
$a_T(\omega)\approx a_T^{(0)}(\omega)+a_T^{(1)}(\omega)$, where
\begin{eqnarray}
a^{(0)}_T (\omega)&=&\int_{-\infty}^{\infty}d\omega'B(\omega,\omega')a^{(0)}_T(\omega-\omega')\cr
&&\times\int_{-\infty}^{\infty}d\omega'' \left[a^{(0)}_T (\omega'') a^{(0)+}_T (\omega'' -\omega')+a^{(0)+}_T(\omega'') a^{(0)}_T (\omega''+\omega')\right]+C(\omega)
\label{noiseeq}
\end{eqnarray}
and
\begin{eqnarray}
a_T^{(1)}(\omega)&=&\int_{-\infty}^{\infty}d\omega'a^{(0)}_T(\omega-\omega')A(\omega,\omega')+\int_{-\infty}^{\infty}d\omega'B(\omega,\omega')a^{(1)}_T(\omega-\omega')\cr
&&\times\int_{-\infty}^{\infty}d\omega'' \left[a^{(0)}_T (\omega'') a^{(0)+}_T(\omega'' -\omega')+a^{(0)+}_T (\omega'') a^{(0)}_T (\omega''+\omega')\right]\cr
&&+\int_{-\infty}^{\infty}d\omega'B(\omega,\omega')a^{(0)}_T(\omega-\omega')\int_{-\infty}^{\infty}d\omega'' \left[a^{(0)}_T (\omega'') a^{(1)+}_T(\omega'' -\omega')\right.\cr
&&\left.+a^{(1)+}_T (\omega'') a^{(0)}_T (\omega''+\omega')+a^{(1)}_T (\omega'') a^{(0)+}_T(\omega'' -\omega')\right.\cr
&&\left.+a^{(0)+}_T (\omega'') a^{(1)}_T (\omega''+\omega')\right].
\label{signaleq}
\end{eqnarray}
Eq.~(\ref{noiseeq}) then yields the detector  noise, while (\ref{signaleq}) yields the detector response to the signal within the linear response approximation. Thus, our approach here is to treat the mechanical oscillator as part of the detector degrees of freedom, with the signal defined as the thermal bath fluctuations and classical external force acting on the oscillator. This is the appropriate viewpoint for force detection. On the other hand, if the focus is on measuring the quantum state of the mechanical oscillator itself, then the oscillator should not be included as part of the detector degrees of freedom. Nevertheless, as we shall later see, the latter viewpoint can be straightforwardly extracted from  the former under not too strong coupling $K_{Tm}$ and pump drive current amplitude $I_0$ conditions. 

\subsection{Semiclassical approximation}
The sequence of solution steps to Eqs.~(\ref{noiseeq}) and (\ref{signaleq})   are in principle as follows: (1) Solve  first equation  (\ref{noiseeq}) for $a_T^{(0)}(\omega)$ in terms of $B(\omega,\omega')$ and $C(\omega)$; (2) Substitute the solution for $a_T^{(0)}(\omega)$ into Eq.~(\ref{signaleq}) for $a_T^{(1)}(\omega)$ and invert this Eq.~(which is linear in $a_T^{(1)}(\omega)$)  to obtain the solution for $a_T^{(1)}(\omega)$ in terms of  $A(\omega,\omega')$, $B(\omega,\omega')$, and $C(\omega)$. 
It is not clear how to carry out these steps in practice, however, since the equations involve products of non-commuting operators. Thus, we must find some way to solve by further approximation. The key observation is that the feedline is in a coherent state, which is classical-like for sufficiently large current amplitude $I_0$ so as to ensure signal amplification. We therefore decompose  $a_T^{(0)}(\omega)$ into a classical, expectation-valued part and quantum, operator-valued fluctuation part, $a_T^{(0)}(\omega)=\left\langle a_T^{(0)}(\omega)\right\rangle + \delta a_T^{(0)}(\omega)$, and subtitute into Eq.~(\ref{noiseeq}) for $a_T^{(0)}(\omega)$, linearizing with respect to the quantum fluctuation $\delta a_T^{(0)}(\omega)$. This gives two equations, one for the expectation value
\begin{eqnarray}
\left\langle a^{(0)}_T (\omega)\right\rangle&=&\int_{-\infty}^{\infty}d\omega'B(\omega,\omega')\left\langle a^{(0)}_T(\omega-\omega')\right\rangle\int_{-\infty}^{\infty}d\omega'' \left[\left\langle a^{(0)}_T (\omega'')\right\rangle \left\langle a^{(0)+}_T (\omega'' -\omega')\right\rangle\right.\cr
&&+\left.\left\langle a^{(0)+}_T(\omega'')\right\rangle \left\langle a^{(0)}_T (\omega''+\omega')\right\rangle\right]+\left\langle C(\omega)\right\rangle
\label{aTaverageeq}
\end{eqnarray}
and the other for the quantum fluctuation:
\begin{eqnarray}
\delta a^{(0)}_T (\omega)&=&\int_{-\infty}^{\infty}d\omega'B(\omega,\omega')\delta a^{(0)}_T(\omega-\omega')\int_{-\infty}^{\infty}d\omega'' \left[\left\langle a^{(0)}_T (\omega'')\right\rangle \left\langle a^{(0)+}_T (\omega'' -\omega')\right\rangle\right.\cr
&&+\left.\left\langle a^{(0)+}_T(\omega'')\right\rangle \left\langle a^{(0)}_T (\omega''+\omega')\right\rangle\right]+\int_{-\infty}^{\infty}d\omega'B(\omega,\omega')\left\langle a^{(0)}_T(\omega-\omega')\right\rangle\cr
&&\times\int_{-\infty}^{\infty}d\omega'' \left[\delta a^{(0)}_T (\omega'') \left\langle a^{(0)+}_T (\omega'' -\omega')\right\rangle+\left\langle a^{(0)}_T (\omega'')\right\rangle \delta a^{(0)+}_T (\omega'' -\omega')\right.\cr
&&\left.+\delta a^{(0)+}_T(\omega'') \left\langle a^{(0)}_T (\omega''+\omega')\right\rangle+\left\langle a^{(0)+}_T(\omega'')\right\rangle \delta a^{(0)}_T (\omega''+\omega')\right]+\delta C(\omega).
\label{aTflucteq}
\end{eqnarray}
Eq.~(\ref{signaleq}) for $a_T^{(1)}(\omega)$ is approximated by replacing $a_T^{(0)}(\omega)$ with its expectation value $\left\langle a_T^{(0)}(\omega)\right\rangle$, i.e., we drop the quantum fluctuation part $\delta a_T^{(0)}(\omega)$. This is because  Eq.~(\ref{signaleq}) already depends linearly on the quantum fluctuating signal term $A(\omega,\omega')$, which we of course want to keep. Dropping the $\delta a_T^{(0)}(\omega)$ contribution to Eq.~(\ref{signaleq}) amounts to neglecting multiplicative detector noise, which is reasonable given that we are concerned with large signal amplification.

\subsection{Complete solution to detector signal response and noise}
The sequence of solutions steps are therefore in practice as follows: (1) Solve Eq.~(\ref{aTaverageeq}) first for $\left\langle a_T^{(0)}(\omega)\right\rangle$; (2) Substitute this solution into Eq.~(\ref{signaleq}) for $a_T^{(1)}(\omega)$ and invert; (3) Substitute the solution for $\left\langle a_T^{(0)}(\omega)\right\rangle$ into the Eq.~(\ref{aTflucteq}) for  $\delta a^{(0)}_T (\omega)$ and invert; (4) Use these solutions for $a_T^{(1)}(\omega)$ and  $\delta a^{(0)}_T (\omega)$  to determine the detector signal and noise terms, respectively. Beginning with step (1), we have
\begin{equation}
\left\langle C(\omega)\right\rangle=-\frac{i\sqrt{2\gamma_{pT}}e^{i\phi_{pT}}}{\gamma_T-i\Delta\omega}\left\langle a_p^{\mathrm{in}}(\omega)\right\rangle
=\frac{i\sqrt{2\gamma_{pT}}e^{i\phi_{pT}}}{\gamma_T-i\Delta\omega}\cdot I_0\sqrt{\frac{Z_p T_M^2}{2\hbar\omega}}e^{-(\omega-\omega_p)^2 T_M^2/2},
\label{Cdefeq}
\end{equation}
where $\Delta\omega=\omega_p-\omega_T$ is the detuning frequency (not to be confused with the bandwidth $\delta\omega$) and note $\left\langle a_e^{\mathrm{in}}(\omega)\right\rangle=0$ (recall, we assume the transmission line resonant frequency $\omega_T$ mode is in the vacuum state).
Given that $T_M$ is the longest timescale in the system dynamics, $\left\langle C(\omega)\right\rangle$ is sharply peaked about the frequency $\omega_p$ and we will therefore approximate the exponential with a delta function: $\left\langle C(\omega)\right\rangle=c\delta(\omega-\omega_p)$,
where
\begin{equation}
c=\frac{i\sqrt{2\pi}e^{i\phi_{pT}}}{\gamma_T-i\Delta\omega}\sqrt{\frac{I_0^2 Z_p \gamma_{pT}}{\hbar\omega_p}}.
\label{ceq}
\end{equation}
Considering for the moment an iterative solution to  Eq.~(\ref{aTaverageeq}) for  $\langle a^{(0)}_T (\omega)\rangle$, we see that $\langle a^{(0)}_T (\omega)\rangle$ must also have the form of a delta function peaked at $\omega_p$:  $\langle a^{(0)}_T (\omega)\rangle=\chi\delta(\omega-\omega_p)$. Substituting this ansatz into  Eq.~(\ref{aTaverageeq}), we obtain the following equation for $\chi$:
\begin{equation}
\chi=2\chi\left|\chi\right|^2 B(\omega_p,0)+c.
\label{chieq}
\end{equation}
This equation has a rather involved analytical solution. For sufficiently large  $\left|c\right|^2 \left|B(\omega_p,0)\right|$ the response can become bistable (i.e., two locally stable solutions for $\chi$). This region will not be discussed in the present paper, however. When we consider actual device parameters later in Sec.~\ref{results}, we will assume sufficiently small drive such that $\chi\approx c$, allowing much simpler analytical expressions to be written down for the detector response.

Proceeding now to step (2), we substitute the expectation value  $\left\langle a^{(0)}_T (\omega)\right\rangle=\chi\delta(\omega-\omega_p)$ for the operator $a^{(0)}_T (\omega)$ into Eq.~(\ref{signaleq}) for $a^{(1)}_T(\omega)$. Carrying out the integrals, we obtain
\begin{eqnarray}
&&\left\{1-2\left|\chi\right|^2 \left[B(\omega,0)+B(\omega,\omega-\omega_p)\right]\right\}a_T^{(1)}(\omega)-2\chi^2 B(\omega,\omega-\omega_p)a^{(1)+}_T (2\omega_p-\omega)\cr
&&=\chi A(\omega,\omega-\omega_p).
\label{aTeq}
\end{eqnarray}
Before we can invert to obtain $a^{(1)}_T(\omega)$, we  require a second linearly independent equation also involving $a^{(1)+}_T (2\omega_p-\omega)$ and $a^{(1)}_T(\omega)$. This equation can be obtained by replacing $\omega$ with $2\omega_p-\omega$ in Eq.~(\ref{aTeq}) and then taking the adjoint:
\begin{eqnarray}
&&\left\{1+2\left|\chi\right|^2 \left[B(\omega-2\Delta\omega,0)+B(\omega-2\Delta\omega,\omega-\omega_p)\right]\right\}a^{(1)+}_T(2\omega_p-\omega)\cr
&&+2\chi^{*2} B(\omega-2\Delta\omega,\omega-\omega_p)a^{(1)}_T (\omega)=-\chi^* A(\omega-2\Delta\omega,\omega-\omega_p),
\label{aTadjeq}
\end{eqnarray}
where we have used the identities $A^+(2\omega_p-\omega,\omega_p-\omega)=-A(\omega-\Delta\omega,\omega-\omega_p)$, $B^*(2\omega_p-\omega,\omega_p-\omega)=-B(\omega-2\Delta\omega,\omega-\omega_p)$, and $B^*(2\omega_p-\omega,0)=-B(\omega-2\Delta\omega,0)$. Inverting, we obtain
\begin{equation}
a_T^{(1)}(\omega)=\alpha_1(\omega)A(\omega,\omega-\omega_p)+\alpha_2(\omega) A(\omega-2\Delta\omega,\omega-\omega_p),
\label{aTsolneq}
\end{equation}
where
\begin{equation}
\alpha_1(\omega)=D(\omega)^{-1}{\left\{1+2\left|\chi\right|^2 \left[ B(\omega-2\Delta\omega,0)+B(\omega-2\Delta\omega,\omega-\omega_p)\right]\right\}\chi}
\label{alphaeq}
\end{equation} and
\begin{equation}
\alpha_2(\omega)=-2 D(\omega)^{-1}\left|\chi\right|^2 B(\omega,\omega-\omega_p)\chi,
\label{alpha2eq}
\end{equation}
with determinant
\begin{eqnarray}
D(\omega)&=& \left\{1-2\left|\chi\right|^2 \left[B(\omega,0)+B(\omega,\omega-\omega_p)\right]\right\} \cr
&&\times\left\{1+2\left|\chi\right|^2 \left[B(\omega-2 \Delta\omega,0)+B(\omega-2 \Delta\omega,\omega-\omega_p)\right]\right\}\cr
&&+4\left|\chi\right|^4 B(\omega,\omega-\omega_p) B(\omega-2\Delta\omega,\omega-\omega_p).
\label{Deteq}
\end{eqnarray}

Moving on now to step (3), we substitute  the expectation value  $\langle a^{(0)}_T (\omega)\rangle=\chi\delta(\omega-\omega_p)$  into Eq.~(\ref{aTflucteq}) for $\delta a^{(0)}_T(\omega)$ and carry out the integrals to obtain:
\begin{eqnarray}
&&\left\{1-2\left|\chi\right|^2 \left[B(\omega,0)+B(\omega,\omega-\omega_p)\right]\right\}\delta a_T^{(0)}(\omega)-2\chi^2 B(\omega,\omega-\omega_p)\delta a^{(0)+}_T (2\omega_p-\omega)\cr
&&=\delta C(\omega).
\label{deltaaeq}
\end{eqnarray}
Replacing $\omega$ with $2\omega_p-\omega$ in Eq.~(\ref{deltaaeq}) and then taking the adjoint:\begin{eqnarray}
&&\left\{1+2\left|\chi\right|^2 \left[B(\omega-2\Delta\omega,0)+B(\omega-2\Delta\omega,\omega-\omega_p)\right]\right\}\delta a^{(0)+}_T(2\omega_p-\omega)\cr
&&+2\chi^{*2} B(\omega-2\Delta\omega,\omega-\omega_p)\delta a^{(0)}_T (\omega)=\delta C^+ (2\omega_p-\omega).
\label{deltaaadjeq}
\end{eqnarray}
Inverting Eqs.~(\ref{deltaaeq}) and (\ref{deltaaadjeq}), we obtain
\begin{equation}
\delta a_T^{(0)}(\omega)=\beta_1(\omega)\delta C(\omega)+\beta_2(\omega)\delta C^+ (2\omega_p-\omega),
\label{deltaasolneq}
\end{equation}
where
\begin{equation}
\beta_1(\omega)=D(\omega)^{-1}{\left\{1+2\left|\chi\right|^2 \left[ B(\omega-2\Delta\omega,0)+B(\omega-2\Delta\omega,\omega-\omega_p)\right]\right\}} 
\label{betaeq}
\end{equation}
and
\begin{equation}
\beta_2(\omega)=2 D(\omega)^{-1}\chi^2 B(\omega,\omega-\omega_p)
\label{beta2eq}
\end{equation}

We are now ready to carry out step (4). To obtain the detector response, we substitute into  expression (\ref{outcurrentsquaredeq}) for $\overline{\left\langle\left[\delta I^{\mathrm{out}}\right]^2\right\rangle}$ the linear response approximation to the `out' probe operator [see Eq.~(\ref{inoutidentity2eq})]:
\begin{equation}
a_p^{\mathrm{out}}(\omega)=\left[-i\sqrt{2\gamma_{pT}}e^{-i\phi_{pT}} a_T^{(1)}(\omega)\right]+\left[-i\sqrt{2\gamma_{pT}}e^{-i\phi_{pT}}\delta a_T^{(0)}(\omega)+\delta a_p^{\mathrm{in}}(\omega)\right].
\label{linearinouteq}
\end{equation}
The first square-bracketed term will give the signal contribution to the detector response, while the second bracketed term gives the noise contribution. Note that the average values $\left\langle a_T^{(0)}(\omega)\right\rangle$ and $\left\langle a_p^{\mathrm{in}}(\omega)\right\rangle$ are not required in the noise term since they give negligible contribution in the signal bandwidths of interest centered at $\omega_s=\omega_p\pm\omega_m$. Substituting in the signal part of $a_p^{\mathrm{out}}(\omega)$, we obtain after some algebra:
\begin{eqnarray}
&&\left.\overline{\left\langle\left[\delta I^{\mathrm{out}}\left(\omega_s,\delta\omega\right)\right]^2\right\rangle}\right|_{\mathrm{signal}}=\cr
&& \left(\frac{I_0 K_{T m} \omega_T}{\gamma_{T}}\right)^2\frac{\gamma^2_{pT}}{\gamma^2_T+\Delta\omega^2} \int_{\omega_s-\delta\omega/2}^{\omega_s+\delta\omega/2}\frac{d\omega}{2\pi}\left[\frac{\omega}{\omega_p}\frac{\gamma_T^2}{(\omega-\omega_p+\Delta\omega)^2+\gamma_T^2}\right]\cr
&&\times\left|\frac{\alpha_1(\omega)}{c}+\frac{\alpha_2(\omega)}{c}\left(\frac{\omega-\omega_p+\Delta\omega+i\gamma_T}{\omega-\omega_p-\Delta\omega+i\gamma_T}\right)\right|^2\cr
&&\times\left\{\frac{2\gamma_{bm}}{(\omega-\omega_p-\omega_m)^2 +\gamma_{bm}^2}\left[2 n(\omega-\omega_p)+1\right]+\frac{2\gamma_{bm}}{(\omega_p-\omega-\omega_m)^2 +\gamma_{bm}^2}\left[2 n(\omega_p-\omega)+1\right]\right\}\cr
&&+\left(\frac{I_0 K_{T m} \omega_T}{\gamma_{T}}\right)^2\frac{\gamma^2_{pT}}{\gamma^2_T+\Delta\omega^2} \frac{1}{2 m\hbar\omega_m \gamma_{bm}}\int_{\omega_s-\delta\omega/2}^{\omega_s+\delta\omega/2}\frac{d\omega d\omega'}{2\pi}\left[\frac{\omega}{\omega_p}\frac{\gamma_T^2}{(\omega-\omega_p+\Delta\omega)^2+\gamma_T^2}\right]\cr
&&\times\left|\frac{\alpha_1(\omega)}{c}+\frac{\alpha_2(\omega)}{c}\left(\frac{\omega-\omega_p+\Delta\omega+i\gamma_T}{\omega-\omega_p-\Delta\omega+i\gamma_T}\right)\right|^2\cr
&&\times\frac{\sin\left[(\omega-\omega')T_M/2\right]}{\left(\omega-\omega'\right) T_M/2}\left\{\frac{2\gamma_{bm}}{(\omega-\omega_p-\omega_m)^2 +\gamma_{bm}^2}F_{\mathrm{ext}}(\omega-\omega_p)F^*_{\mathrm{ext}}(\omega'-\omega_p)\right.\cr
&&\left.+\frac{2\gamma_{bm}}{(\omega_p-\omega-\omega_m)^2 +\gamma_{bm}^2}F_{\mathrm{ext}}(\omega_p-\omega)F^*_{\mathrm{ext}}(\omega_p-\omega')\right\},
\label{completesignaleq}
\end{eqnarray} 
where $n(\omega)=\left(e^{\hbar\omega/k_B T}-1\right)^{-1}$ is the Bose-Einstein thermal occupation number average for bath mode $\omega$.
The signal part of the detector response comprises a thermal component and a classical force component. In the limit of weak coupling $K_{Tm}\rightarrow 0$ and or small drive current amplitude $I_0\rightarrow 0$, we have $\alpha_1(\omega)/c\rightarrow 1$,  $\alpha_2(\omega)/c\rightarrow 0$ and we note that the frequency resolved detector response has the form of two Lorentzians centered at $\omega_p\pm\omega_m$. The resulting expression for the detector response coincides with an $O(K_{Tm}^2)$ perturbative solution to the detector response~(\ref{outcurrentsquaredeq}) via the linear response Eqs.~(\ref{noiseeq}) and (\ref{signaleq}) (but no semiclassical approximation).  However, as shall be described in Sec.~\ref{results}, when the current drive is not small and or coupling is not weak, then the $\alpha_i$ terms will modify this simple form, at the next level of approximation renormalizing the Lorentzians, i.e., shifting their location and changing their width.      

Substituting in the noise part of $a_p^{\mathrm{out}}(\omega)$, we obtain after some algebra:
\begin{eqnarray}
&&\left.\overline{\left\langle\left[\delta I^{\mathrm{out}}\left(\omega_s,\delta\omega\right)\right]^2\right\rangle}\right|_{\mathrm{noise}}=Z_p^{-1}\int_{\omega_s-\delta\omega/2}^{\omega_s+\delta\omega/2}\frac{d\omega}{2\pi}\hbar\omega\frac{2\gamma_T\gamma_{pT}}{(\omega-\omega_p+\Delta\omega)^2+\gamma_T^2}\cr
&&\times\left\{\left|\beta_1(\omega)\right|^2 +\frac{(\omega-\omega_p+\Delta\omega)^2+\gamma^2_T}{(\omega-\omega_p-\Delta\omega)^2+\gamma^2_T}\left|\beta_2(\omega)\right|^2-\mathrm{Re}\left[\beta_1(\omega)\right] +\frac{(\omega-\omega_p +\Delta\omega)}{\gamma_T}\mathrm{Im}\left[\beta_1(\omega)\right]\right\}\cr
&&+Z_p^{-1}\frac{\hbar\omega_s}{2}\frac{\delta\omega}{2\pi}.
\label{completenoiseeq}
\end{eqnarray}
The noise part of the detector response  comprises a back reaction component (the integral term) where transmission line noise drives the mechanical oscillator via the SQUID coupling, and a component that is added at the output due to zero-point fluctuations in the probe line. While not as obvious given the form of Eq.~(\ref{completenoiseeq}), one may again verify (see Sec.~\ref{results}) that the detector back reaction on the mechanical oscillator takes the form of two Lorentzians centered at $\omega_p\pm\omega_m$  in the weak coupling and or weak current drive limit, coinciding with an $O(K_{Tm}^2)$ perturbative calculation.  

Eqs.~(\ref{completesignaleq}) and (\ref{completenoiseeq}) are the main results of the paper, their sum giving the net output mean-squared current.

\subsection{Quantum bound on noise}
As articulated by Caves,\cite{caves} the fact that the `in' and `out' operators  satisfy canonical commutation relations places a lower, quantum limit on the noise contribution to the detector response, Eq.~(\ref{completenoiseeq}). We now derive this quantum limit. First write the `out' operator (\ref{linearinouteq}) as
\begin{equation}
a_p^{\mathrm{out}}(\omega)=-i\sqrt{2\gamma_{pT}}e^{-i\phi_{pT}}a_T^{(1)}(\omega)+N(\omega),
\label{linearinout2eq}
\end{equation}
where $N(\omega)=-i\sqrt{2\gamma_{pT}}e^{-i\phi_{pT}}\delta a_T^{(0)}(\omega)+\delta a_p^{\mathrm{in}}(\omega)$ is the noise part.
Taking commutators, we have the following identity relating the noise and signal operator terms:
\begin{equation}
\left[N(\omega),N^+(\omega')\right]=\delta(\omega-\omega')-2\gamma_{pT}\left[a_T^{(1)}(\omega),a_T^{(1)+}(\omega')\right].
\label{commutatoreq}
\end{equation}
Now, from the Heisenberg Uncertainty Principle, one can derive the following general inequality:
\begin{equation}
\left\langle N[f]N^+[f]+N^+[f]N[f]\right\rangle\geq\left|\left\langle\left[N[f],N^+[f]\right]\right\rangle\right|,
\label{hupeq}
\end{equation}
where  $N[f]=\int_0^{\infty}d\omega f(\omega)N(\omega)$ and $f(\omega)$ is an arbitrary function. 
Inserting the commutator identity~(\ref{commutatoreq}), Eq.~(\ref{hupeq}) becomes
\begin{equation}
\left\langle N[f]N^+[f]+N^+[f]N[f]\right\rangle\geq\left|\int_0^{\infty}d\omega \left|f(\omega)\right|^2 -2\gamma_{pT}\left\langle\left[a_T^{(1)}[f],a_T^{(1)+}[f]\right]\right\rangle\right|.
\end{equation}
Choosing the `filter' function  $f(\omega)=\omega\Theta(\omega-\omega_s+\delta\omega/2)\Theta(\omega_s+\delta\omega/2-\omega)$ and evaluating the commutator, we obtain the following lower bound on the detector noise:
\begin{eqnarray}
&&\left.\overline{\left\langle\left[\delta I^{\mathrm{out}}\left(\omega_s,\delta\omega\right)\right]^2\right\rangle}\right|_{\mathrm{noise}}\geq\cr
&&\left|Z_p^{-1}\frac{\hbar\omega_s}{2}\frac{\delta\omega}{2\pi}-\left(\frac{I_0 K_{T m} \omega_T}{\gamma_{T}}\right)^2\frac{\gamma^2_{pT}}{\gamma^2_T+\Delta\omega^2}\right.
 \int_{\omega_s-\delta\omega/2}^{\omega_s+\delta\omega/2}\frac{d\omega}{2\pi}\left[\frac{\omega}{\omega_p}\frac{\gamma_T^2}{(\omega-\omega_p+\Delta\omega)^2+\gamma_T^2}\right]\cr
&&\times\left|\frac{\alpha_1(\omega)}{c}+\frac{\alpha_2(\omega)}{c}\left(\frac{\omega-\omega_p+\Delta\omega+i\gamma_T}{\omega-\omega_p-\Delta\omega+i\gamma_T}\right)\right|^2\cr
&&\times\left.\left[\frac{2\gamma_{bm}}{(\omega-\omega_p-\omega_m)^2 +\gamma_{bm}^2}-\frac{2\gamma_{bm}}{(\omega_p-\omega-\omega_m)^2 +\gamma_{bm}^2}\right]\right|.
\label{noiseboundeq}
\end{eqnarray}
In the next section we will address the extent to which the detector noise can approach the quantum bound on the right hand side of Eq.~(\ref{noiseboundeq}), depending on the current drive amplitude $I_0$ and other detector parameters.

\section{Results}
\label{results}

\subsection{Analytical approximations}
To gain a better understanding of the detector response, we now provide  analytical approximations to Eqs.~(\ref{completesignaleq}) and (\ref{completenoiseeq}) that are valid under the condition
$\left|c\right|^2 \left|B(\omega_p,0)\right|\ll 1$ such that $\chi\approx c$  [see Eq.~(\ref{chieq})], i.e., the expectation value $\left\langle a_T^{(0)}(\omega)\right\rangle$ for the transmission line depends approximately only on the pump/probe feedline state and not on the mechanical oscillator state. Explicitly, this condition reads:
\begin{equation}
\frac{2 I_0^2 Z_p K_{Tm}^2  \omega_T \gamma_{pT}}{\hbar\omega_m \left(\gamma_T^2 +\Delta\omega^2\right)^{3/2}}\ll1,
\label{analyticcondeq}
\end{equation}
placing an upper limit on $I_0$ and $K_{Tm}$ for the validity of this approximation. We also assume  that the mechanical and transmission line mode frequencies are widely separated: $\omega_m\ll\omega_T$, and with small damping rates: $\gamma_{bm}\ll\omega_m$, $\gamma_T\ll\omega_T$. We do not restrict  the relative magnitudes of $\omega_m$ and $\gamma_T$, however.  A simple picture emerges in which the detector back reaction  `renormalizes' the mechanical oscillator  frequency and damping rate:
$\omega_m\rightarrow R_{\omega}\omega_m$ and $\gamma_{bm}\rightarrow R_{\gamma}\gamma_{bm}$, where
\begin{equation}
R_{\omega}\omega_m=\omega_m+\left(\Delta\omega+\frac{|c|^2 \omega_T^2 K_{Tm}^2}{\pi\omega_m}\right)\frac{{|c|^2 \omega_T^2 K_{Tm}^2}\left[\gamma_T^2+\Delta\omega^2-\omega_m^2\right]}{\pi\left[\gamma_T^2+\left(\Delta\omega+\omega_m\right)^2\right]\left[\gamma_T^2+\left(\Delta\omega-\omega_m\right)^2\right]}
\label{Romegaeq}
\end{equation}
and
\begin{equation}
R_{\gamma}\gamma_{bm}=\gamma_{bm}-\left(\Delta\omega+\frac{|c|^2 \omega_T^2 K_{Tm}^2}{\pi\omega_m}\right)\frac{{2|c|^2\omega_T^2 K_{Tm}^2\omega_m\gamma_T}}{\pi\left[\gamma_T^2+\left(\Delta\omega+\omega_m\right)^2\right]\left[\gamma_T^2+\left(\Delta\omega-\omega_m\right)^2\right]},
\label{Rgammaeq} 
\end{equation}
where $c$ is defined in Eq.~(\ref{ceq}).
With the measurement filter bandwidth centered at either of $\omega_s=\omega_p\pm R_{\omega}\omega_m$, the approximation to Eq.~(\ref{completesignaleq}) for the signal response is (with the classical force term omitted):
\begin{eqnarray}
&&\left.\overline{\left\langle\left[\delta I^{\mathrm{out}}\left(\omega_s=\omega_p\pm R_{\omega}\omega_m,\delta\omega\right)\right]^2\right\rangle}\right|_{\mathrm{signal}}= \left(\frac{I_0 K_{T m} \omega_T}{\gamma_{T}}\right)^2\frac{\gamma^2_{pT}}{\gamma^2_T+\Delta\omega^2}\frac{\gamma_T^2}{\gamma_T^2+\left(\Delta\omega\pm\omega_m\right)^2} \cr
&&\times\int_{\omega_s-\delta\omega/2}^{\omega_s+\delta\omega/2}\frac{d\omega}{2\pi}\frac{2\gamma_{bm}}{(\omega-\omega_p\mp R_{\omega}\omega_m)^2 +\left(R_{\gamma}\gamma_{bm}\right)^2}\left[2 n(R_{\omega}\omega_m)+1\right].
\label{approxsignaleq}
\end{eqnarray}
When there is a classical force acting on the mechanical oscillator, we must add to Eq.~(\ref{approxsignaleq}) the term
\begin{eqnarray} 
&&\left(\frac{I_0 K_{T m} \omega_T}{\gamma_{T}}\right)^2\frac{\gamma^2_{pT}}{\gamma^2_T+\Delta\omega^2} \frac{1}{2 m\hbar\omega_m \gamma_{bm}}\int_{\omega_s-\delta\omega/2}^{\omega_s+\delta\omega/2}\frac{d\omega d\omega'}{2\pi}\frac{\gamma_T^2}{\left(\omega-\omega_p+\Delta\omega\right)^2+\gamma_T^2}\cr
&&\times\frac{\sin\left[(\omega-\omega')T_M/2\right]}{\left(\omega-\omega'\right) T_M/2}\left\{\frac{2\gamma_{bm}}{(\omega-\omega_p-R_{\omega}\omega_m)^2 +\left(R_{\gamma}\gamma_{bm}\right)^2}F_{\mathrm{ext}}(\omega-\omega_p)F^*_{\mathrm{ext}}(\omega'-\omega_p)\right.\cr
&&\left.+\frac{2\gamma_{bm}}{(\omega_p-\omega-R_{\omega}\omega_m)^2 +\left(R_{\gamma}\gamma_{bm}\right)^2}F_{\mathrm{ext}}(\omega_p-\omega)F^*_{\mathrm{ext}}(\omega_p-\omega')\right\}.
\label{approxforceeq}
\end{eqnarray} 
The approximation to Eq.~(\ref{completenoiseeq}) for the detector noise is 
\begin{eqnarray}
&&\left.\overline{\left\langle\left[\delta I^{\mathrm{out}}\left(\omega_s=\omega_p\pm R_{\omega}\omega_m,\delta\omega\right)\right]^2\right\rangle}\right|_{\mathrm{noise}}= \left(\frac{I_0 K_{T m} \omega_T}{\gamma_{T}}\right)^2\frac{\gamma^2_{pT}}{\gamma^2_T+\Delta\omega^2}\frac{\gamma_T^2}{\gamma_T^2+\left(\Delta\omega\pm\omega_m\right)^2} \cr
&&\times\int_{\omega_s-\delta\omega/2}^{\omega_s+\delta\omega/2}\frac{d\omega}{2\pi}\frac{2\gamma_{bm}}{(\omega-\omega_p\mp R_{\omega}\omega_m)^2 +\left(R_{\gamma}\gamma_{bm}\right)^2}N_{\pm} +Z_p^{-1}\frac{\hbar\omega_s}{2}\frac{\delta\omega}{2\pi},
\label{approxnoiseeq}
\end{eqnarray}
where the back reaction noise parameter is
\begin{equation}
N_{\pm}=\frac{|c|^2 K_{Tm}^2\omega^2_T\gamma_{T}}{\pi\gamma_{bm}\left[\gamma_T^2+\left(\Delta\omega\mp\omega_m\right)^2\right]}\mp 1=\frac{2I_0^2Z_pK_{Tm}^2\omega_T\gamma_{T}\gamma_{pT}}{\hbar\gamma_{bm}\left[\gamma_T^2+\Delta\omega^2\right]\left[\gamma_T^2+\left(\Delta\omega\mp\omega_m\right)^2\right]}\mp 1.
\label{backnoiseeq}
\end{equation}
The $\mp1$ term in the back reaction noise parameter depends on whether the filter is centered at $\omega_s=\omega_p+\omega_m$ or $\omega_s=\omega_p-\omega_m$ and corresponds respectively to  `phase preserving'  or  `phase conjugating' detection as discussed in Caves.\cite{caves} In the limit $I_0\rightarrow 0$ and or $K_{Tm}\rightarrow 0$, we see from Eqs.~(\ref{approxsignaleq}), (\ref{approxnoiseeq}), and (\ref{backnoiseeq}) that the back reaction noise amounts to doubling the oscillator quantum zero-point motion signal in the phase conjugating case, while the back reaction noise exactly cancels the quantum zero-point motion signal in the phase preserving case. In both cases, the  noise coincides with the lower quantum bound~(\ref{noiseboundeq}). However, in this small drive/coupling limit, we do not have a detector or amplifier but rather an attenuator, which is  of only academic interest to us. 

Comparing the detector response~(\ref{approxsignaleq}) and back reaction part of Eq.~(\ref{approxnoiseeq}), we see that the mechanical oscillator behaves in the steady state as if in contact with a thermal bath.\cite{mozyrskyprl02,armourprb04,clerkprb04,blencowecontempphys05,clerknjp05,blencowenjp05,marquardtprpt07} The back reaction of the detector on the mechanical oscillator is effectively that of a thermal bath with damping rate $\gamma_{\mathrm{back}}=\gamma_{bm} (R_{\gamma}-1)$ and effective thermal average occupation number $n_{\mathrm{back}}$ defined as follows:
\begin{equation}
\gamma_{\mathrm{back}} (2n^{\pm}_{\mathrm{back}}+1)=\gamma_{bm} N_{\pm}.
\label{ndet0eq}
\end{equation}
Thus,
\begin{equation}
n^{\pm}_{\mathrm{back}}=\left(R_{\gamma}-1\right)^{-1} \frac{1}{2}N_{\pm}-\frac{1}{2}.
\label{ndeteq}
\end{equation}
The failure to approach the lower quantum bound (\ref{noiseboundeq}) when $N_{\pm}\gg 1$ then translates into having $(2n^{\pm}_{\mathrm{back}}+1)\gamma_{\mathrm{back}}/\gamma_{bm}\gg 1$. Thus, to get close to the bound, we necessarily require $\gamma_{\mathrm{back}}\ll\gamma_{bm}$;\cite{clerkprb04} the
back reaction occupation number  $n^{\pm}_{\mathrm{back}}$ does not have to be small.
With the mechanical oscillator also in thermal contact with its external bath, the net damping rate of the oscillator is $\gamma_{\mathrm{net}}=\gamma_{bm}+\gamma_{\mathrm{back}}=R_{\gamma}\gamma_{bm}$ and the  net, effective thermal average occupation number $n_{\mathrm{net}}$ of the oscillator is defined as follows:
\begin{equation}
\gamma_{\mathrm{net}}\left(2n^{\pm}_{\mathrm{net}}+1\right)=\gamma_{bm} \left[2n(R_{\omega}\omega_m)+1\right]+\gamma_{\mathrm{back}} \left(2n^{\pm}_{\mathrm{back}}+1\right).
\end{equation}
Thus,
\begin{equation}
n^{\pm}_{\mathrm{net}}=R_{\gamma}^{-1}\left[n(R_{\omega}\omega_m)+\frac{1}{2}+\frac{1}{2}N_{\pm}\right]-\frac{1}{2}.
\label{nneteq}
\end{equation}

From Eq.~(\ref{Rgammaeq}), we see that depending on the detuning parameter $\Delta\omega=\omega_p-\omega_T$, the damping rate of the oscillator due to the detector back reaction can be either negative or positive. Specifically, positive damping requires the following condition on the detuning parameter:
\begin{equation}
\Delta\omega<-\frac{|c|^2 \omega_T^2 K_{Tm}^2}{\pi\omega_m}=-\frac{2 I_0^2 Z_p K_{Tm}^2  \omega_T \gamma_{pT}}{\hbar\omega_m \left(\gamma_T^2 +\Delta\omega^2\right)}.
\label{posdampeq}
\end{equation}

\subsection{Displacement sensitivity}
 In the absence of a classical force acting on the mechanical oscillator, from Eq.~(\ref{approxsignaleq}) the mechanical oscillator thermal noise displacement signal spectral density takes the familiar Lorentzian form:
\begin{equation}
\left. S_x (\omega)\right|_{\mathrm{signal}}=\frac{2 R_{\gamma}\gamma_{bm}}{\left(\omega-\omega_p\mp R_{\omega}\omega_m\right)^2+\left(R_{\gamma}\gamma_{bm}\right)^2}\frac{\hbar}{2 m R_{\omega}\omega_m} \left[2n\left(R_{\omega}\omega_m\right)+1\right].
\label{noisesignaleq}
\end{equation} 
In order to be able to resolve this mechanical signal,  the detector noise (\ref{approxnoiseeq}) referred to the mechanical oscillator input must be smaller than (\ref{noisesignaleq}). The detector noise spectral density at the input is
\begin{eqnarray}
\left. S_x \left(\omega=\omega_p\pm R_{\omega}\omega_m\right)\right|_{\mathrm{noise}}&=&
\left\{\frac{2}{R_{\gamma}\gamma_{bm}}\left[\mp 1+\frac{|c|^2 K_{Tm}^2 \omega_T^2\gamma_T}{\pi\gamma_{bm}\left[\gamma_T^2 +\left(\Delta\omega\mp\omega_m\right)^2\right]}\right]\right.\cr
&&\left.+\frac{2\pi R_{\gamma}\left[\gamma_T^2 +\left(\Delta\omega\pm\omega_m\right)^2\right]}{|c|^2 K_{Tm}^2\omega_T^2\gamma_{pT}}\right\}\frac{\hbar}{2 m R_{\omega}\omega_m},
\label{noisespectrumeq}
\end{eqnarray} 
where  the first term on the right hand side is the back reaction noise acting on the mechanical oscillator and the second term is the output, probe line zero-point noise referred to the input. Note that the noise has been evaluated at $\omega=\omega_p\pm R_{\omega}\omega_m$, the maximum of the back reaction Lorentzian. 

If the detector output is to depend linearly on the mechanical oscillator signal input (i.e., function as a linear amplifier), then back reaction effects must be small. In particular, we require that $\gamma_{\mathrm{back}}\ll\gamma_{bm}$, i.e., $R_{\gamma}\approx 1$. With $|c|$ being proportional $I_0$, we see from Eq.~(\ref{noisespectrumeq}) that increasing the drive current amplitude $I_0$  increases the back reaction noise, but decreases the probe line noise referred to the mechanical oscillator input. Thus, there is an optimum  $I_0$ such that the sum  $\left. S_x\right|_{\mathrm{noise}}$ is a minimum. Making the approximation $R_{\gamma}=1$ and $R_{\omega}=1$ in Eq.~(\ref{noisespectrumeq}) and optimizing with respect to $|c|$, we find
\begin{equation}
\left. S_x \left(\omega=\omega_p\pm R_{\omega}\omega_m\right)\right|_{\mathrm{noise-optimum}}
=\frac{\hbar}{ m \omega_m\gamma_{bm}}\left[\mp 1+2\sqrt{\left(\frac{\gamma_T}{\gamma_{pT}}\right)\frac{\gamma_T^2 +\left(\Delta\omega\pm\omega_m\right)^2}{\gamma_T^2 +\left(\Delta\omega\mp\omega_m\right)^2}}\right].
\label{noiseoptimumeq}
\end{equation}
From Eq.~(\ref{noiseoptimumeq}), we see that the noise is further reduced if (i) the dominant source of transmission line mode dissipation is due to energy loss through the coupled probe (information gathering) line:\cite{clerkprb04} $\gamma_T\approx\gamma_{pT}$; (ii) the detuning frequency is chosen to be $\Delta\omega=\mp\sqrt{\gamma_T^2+\omega_m^2}$, where the minus (plus) sign corresponds to phase preserving (conjugating) detection. With this detuning choice, the condition $R_{\gamma}\approx 1$ requires $\left(\omega_m/\gamma_T\right)^2\ll1$ and so the minimum detector noise is
\begin{equation}
\left. S_x \left(\omega=\omega_p\pm R_{\omega}\omega_m\right)\right|_{\mathrm{noise-optimum}}=
\frac{\hbar}{ m \omega_m\gamma_{bm}}\left[2\mp 1 +O\left((\omega_m/\gamma_T)^2\right)\right],
\label{finalnoiseoptimumeq}
\end{equation}
where in order to determine the $O\left((\omega_m/\gamma_T)^2\right)$ term,  the full form of $R_{\gamma}$ given in Eq.~(\ref{Rgammaeq}) must be used in Eq.~(\ref{noisespectrumeq}) when optimizing.  Comparing with Eq.~(\ref{noisesignaleq}) for the signal noise, we see that to leading order the detector noise effectively doubles the zero-point signal in the phase preserving case. This exceeds the lower bound on the detector noise derived from Eq.~(\ref{noiseboundeq}), which is zero to leading order in the phase preserving case.
\begin{figure}[htbp]
\centering
\epsfig{file=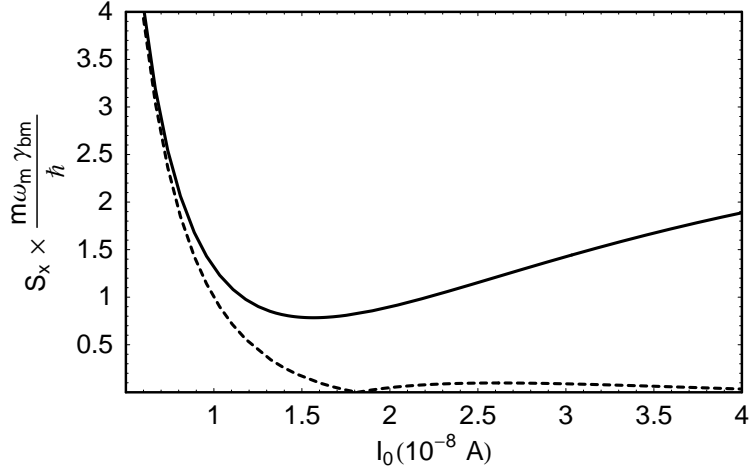,width=4.in} 
\caption{Displacement detector noise spectral density (solid line) and lower bound (dashed line) versus drive current amplitude. The noise densities are evaluated at $\omega=\omega_p+R_{\omega}\omega_m$, corresponding to phase preserving detection.}
\label{displacenoisefig}
\end{figure} 

We now numerically evaluate Eq.~(\ref{noisespectrumeq}) for the detector noise. The feasible example parameter values we use are:\cite{buksprb06}  $B_{\mathrm{ext}}=0.005$~Tesla, $Z_p=50$~Ohms, $\omega_T/2\pi=3 \times 10^9~{\mathrm s}^{-1}$, $Q_T=\omega_T/(2 \gamma_T)=100$, $\gamma_T=9.4\times 10^7~{\mathrm s}^{-1}$,   $l_{\mathrm{osc}}=5~\mu{\mathrm m}$, $\lambda=1$ (geometrical correction factor),  $m=10^{-16}~{\mathrm{kg}}$, $\omega_m=2.5\times 10^7~{\mathrm s}^{-1}$,  and $Q_{bm}=\omega_m/(2\gamma_{bm})=10^3$.  These values give a mechanical oscillator zero-point uncertainty $\Delta x_{zp}=1.45\times 10^{-13}~{\mathrm m}$, a zero-point displacement noise $\hbar/(m\omega_m \gamma_{bm})=3.4\times 10^{-30}~{\mathrm{m}}^2/{\mathrm{Hz}}$, and a dimensionless coupling strength $K_{Tm}=-1.1\times10^{-5}$, where we assume that  in the expression~(\ref{coupling}) for $K_{Tm}$, $\Phi_{\mathrm{ext}}$ can be chosen such that the dimensionless factor $\frac{\Phi_0}{4\pi L_T l I_c}\tan\left(\pi\Phi_{\mathrm{ext}}/\Phi_0\right)\sec\left(\pi\Phi_{\mathrm{ext}}/\Phi_0\right)\approx 1$ (matching condition). We also suppose that $\gamma_T\approx\gamma_{pT}$, i.e., the transmission line mode damping is largely due to the probe line coupling.
 
Fig.~\ref{displacenoisefig} shows $\left. S_x \left(\omega=\omega_p+ R_{\omega}\omega_m\right)\right|_{\mathrm{noise}} \times m\omega_m\gamma_{bm}/\hbar$ and also the lower bound on the detector noise that follows from Eq.~(\ref{noiseboundeq}) for phase preserving detection. Note that the minimum detector noise is approximately $0.8\ \hbar/(m \omega_m\gamma_{bm})$. Thus, for this example, the next-to-leading $O\left((\omega_m/\gamma_T)^2\right)$ term in Eq.~(\ref{finalnoiseoptimumeq}) is approximately $-0.2$. Note also that the detector noise coincides with the lower bound in the small drive limit.

\subsection{Force sensitivity}
Consider a monochromatic classical driving force with frequency $\omega_0\sim R_{\omega}\omega_m$ acting on the oscillator:  $F_{\mathrm{ext}}(\omega)=F_0 \delta(\omega-\omega_0)$. The force signal spectral density is then $\left. S_F(\omega)\right|_{\mathrm{signal}}=F_0^2\delta(\omega-\omega_0)$. For force detection operation, the mechanical oscillator is included as part of the detector degrees of freedom.  From Eqs.~(\ref{approxsignaleq}-\ref{backnoiseeq}), the force noise spectral density evaluated at $\omega=\omega_p\pm\omega_0$ is
\begin{eqnarray}
\left. S_F\left(\omega=\omega_p\pm \omega_0\right)\right|_{\mathrm{noise}}&=&2 m \hbar\omega_m\gamma_{bm}\left\{ 2n\left(\omega_0\right)+1 \mp 1 +\frac{|c|^2 K_{Tm}^2\omega_T^2\gamma_T}{\pi\gamma_{bm}\left[\gamma_T^2+\left(\Delta\omega\mp\omega_m\right)^2\right]}\right.\cr
&&\left.+\frac{\pi\left[\left(\omega_0-R_{\omega}\omega_m\right)^2+\left(R_{\gamma}\gamma_{bm}\right)^2\right] \left[\gamma_T^2+\left(\Delta\omega\pm\omega_m\right)^2\right]}{\gamma_{bm}|c|^2 K_{Tm}^2\omega_T^2\gamma_{pT}}\right\}.
\label{forcenoiseeq}
\end{eqnarray}
Comparing  the displacement noise (\ref{noisespectrumeq}) with the force noise (\ref{forcenoiseeq}), we see that the latter includes the additional $2m\hbar\omega_m\gamma_{bm}\left[2n\left(\omega_0\right)+1\right]$ mechanical quantum thermal displacement noise term. Since the mechanical oscillator forms part of the force detector, it need not necessarily be weakly driven and or weakly coupled to the transmission line;  as explained in Sec.~\ref{linresponse}, the present analysis employs a linear response approximation for force detection, not displacement detection.  Thus, in determining the optimum $I_0$ (and or $K_{Tm}$) and $\Delta\omega$ such that $\left. S_F\right|_{\mathrm{noise}}$ is a minimum, we should not assume \textit{a priori} the restrictions $R_{\gamma},R_{\omega}\approx1$.
\begin{figure}[htbp]
\centering
\epsfig{file=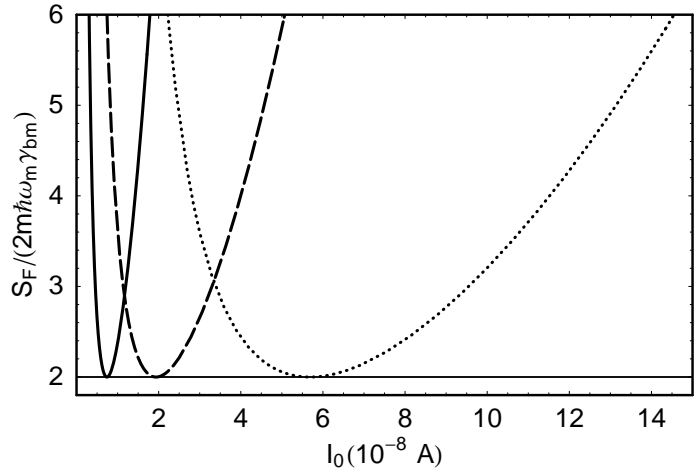,width=4.in} 
\caption{Force detector noise spectral density versus drive current amplitude for detuning $\Delta\omega=0$ (solid line), $\Delta\omega=-5\omega_m$ (dashed line), and $\Delta\omega=-10\omega_m$ (dotted line) . The noise densities are evaluated at $\omega=\omega_p+R_{\omega}\omega_m$, corresponding to phase preserving detection.}
\label{forcenoisefig}
\end{figure}

Fig.~\ref{forcenoisefig} shows the results of numerically evaluating the force noise spectral density given by Eq.~(\ref{forcenoiseeq})  for phase preserving detection ($\omega=\omega_p+\omega_0$) and a range of detuning values. The  same example parameters  are used as in the above displacement sensitivity analysis, with $n(\omega_0)=0$ and $\omega_0 =R_{\omega}\omega_m$. The force noise is expressed in units $2m\hbar\omega_m\gamma_{bm}=6.6\times 10^{-39}~{\mathrm{N}}^2/{\mathrm{Hz}}$. Note that the minimum force noise is exactly 2 in these units, independently of the detuning, with the minimum occuring at larger $I_0$ values as the detuning is made progressively more negative.

\subsection{Back reaction cooling}
From Eq.~(\ref{nneteq}), we see that the net, thermal average occupation number $n_{\mathrm{net}}$ of the mechanical oscillator's fundamental mode decreases as $R_{\gamma}$ increases. Thus, by increasing the drive and or coupling strength such that $\gamma_{\mathrm{back}}\gg\gamma_{bm}$, the mechanical oscillator can be effectively cooled at the expense of increasing its  damping rate.\cite{braginskypla02,martinprb04,wilsonprl04,clerknjp05,blencowenjp05,wineland06,marquardtprpt07,hohbergernature04,naik,harrisrsi07,gigannature06,arcizetnature06,vahalaprl06,corbittprpt06} Consider sufficiently negative detuning such that $-\Delta\omega\gg |c|^2 \omega_T^2 K_{Tm}^2 /(\pi\omega_m)$ [see Eq.~(\ref{posdampeq})]. Substituting  definition~(\ref{Rgammaeq}) for $R_{\gamma}$ and definition~(\ref{backnoiseeq}) for $N_+$ into Eq.~(\ref{nneteq}) and supposing $R_{\gamma}$ is large enough that we can neglect the external damping term $\gamma_{bm}$, we obtain approximately for the phase preserving case:
\begin{equation}
n_{\mathrm{net}}^+ \approx \frac{n\left(R_{\omega}\omega_m\right)}{R_{\gamma}}+n^+_{\mathrm{back}}, 
\label{ncooleq}
\end{equation}
where
\begin{equation}
n^+_{\mathrm{back}}\approx-\frac{\gamma_T^2+\left(\Delta\omega+\omega_m\right)^2}{4\Delta\omega\omega_m}-\frac{1}{2}.
\label{detectorneq}
\end{equation}       
This expression agrees with that derived in Ref.~\onlinecite{marquardtprpt07}, apart from the $1/2$ which is simply due to a small difference in the way we define $n^{\pm}_{\mathrm{back}}$ in Eq.~(\ref{ndet0eq}). Choosing optimum detuning $\Delta\omega=-\sqrt{\gamma_T^2 +\omega_m^2}$  to minimize $n^+_{\mathrm{back}}$ in Eq.~(\ref{detectorneq}), we therefore have
\begin{equation}
n_{\mathrm{net}}^+ \approx\frac{n\left(R_{\omega}\omega_m\right)}{R_{\gamma}}+\frac{1}{2} \sqrt{1+\left(\gamma_T/\omega_m\right)^2}-1.
\label{ncool2eq}
\end{equation}
How much cooling can be achieved depends on (i) how large $R_{\gamma}$ can be,  subject to the above inequality on $-\Delta\omega$; (ii) making the ratio  $\gamma_T/\omega_m$ as small as possible.\cite{marquardtprpt07} 

\begin{figure}[htbp]
\centering
\epsfig{file=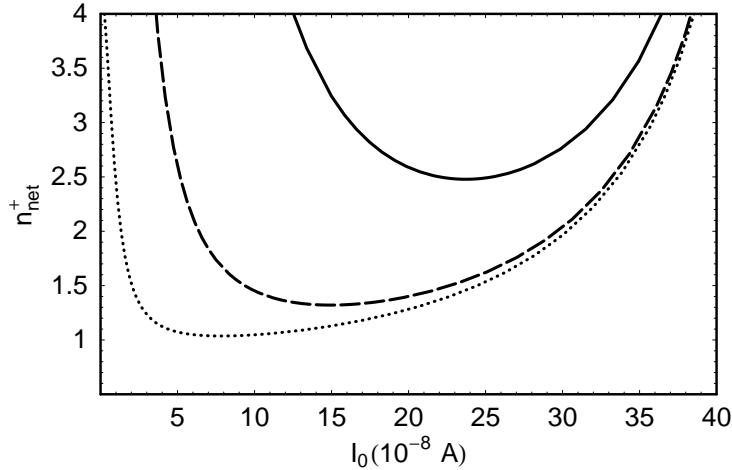,width=4.in} 
\caption{Net effective average occupation number of the mechanical oscillator versus drive current. The solid curve is for external bath temperature $T=100~{\mathrm{mK}}$ $[n(R_{\omega}\omega)=523]$, the dashed curve is for  $T=10~{\mathrm{mK}}$ $[n(R_{\omega}\omega)=52]$, and the dotted curve is for $T=1~{\mathrm{mK}}$ $[n(R_{\omega}\omega)=4.8]$.}  
\label{nnetpfig}
\end{figure}

Using the same example parameter values as above, but taking instead a larger but still realistic  quality factor $Q_{bm}=10^4$ for the mechanical oscillator,\cite{blencowepr04}  the resulting numerically evaluated effective occupation number $n_{\mathrm{net}}^+$ [Eq.~(\ref{nneteq})]  is given in Fig.~\ref{nnetpfig} for a range of external bath occupation numbers $n(\omega_m)$. 
Thus, even for small coupling strengths $K_{Tm}$ and drive current  amplitudes $I_0$, significant cooling of the mechanical oscillator can be achieved. This is in part a consequence of the fact that the quality factor $Q_{bm}$ of the mechanical oscillator when decoupled from the detector  is very large.

\section{Concluding remarks}
\label{conclusion}
In the present paper, we have attempted to give a reasonably comprehensive analysis of the quantum-limited detection sensitivity of a DC SQUID for drive currents well below the Josephson junction critical current $I_c$. In this regime, the  SQUID functions effectively as a mechanical position-dependent inductance element to a good approximation and the resulting closed system Hamiltonian (\ref{hamiltonian3eq})  takes the same form as that for several other types of coupled mechanical resonator-detector resonator systems. Thus, the key derived expressions~(\ref{completesignaleq}) and (\ref{completenoiseeq}) for the detector response and detector noise are of more general application. 

The main approximation made in analyzing the position and force detection sensitivity, as well as back reaction cooling, was to limit the drive current and or coupling strength according to Eq.~(\ref{analyticcondeq}). This allowed us to find much simpler, analytical approximations to the key expressions, in particular Eqs.~(\ref{approxsignaleq}) and (\ref{approxnoiseeq}). The regime of larger drive currents and or coupling strengths which exceed the limit~(\ref{analyticcondeq}) remains to be explored. However, with the SQUID in mind, it is more appropriate to consider larger drive currents in the context of including the non-linear $I/I_c$ corrections to the SQUID effective inductance. This will be the subject of a forthcoming paper.\cite{nation??} 

\section*{Acknowledgements}
We thank A. Armour, A. Rimberg, and P. Nation for helpful discussions. This work was partly supported by the US-Israel Binational
Science Foundation (BSF) and by the National Science Foundation (NSF) under NIRT grant CMS-0404031.  One of the authors (M.P.B.) thanks the Aspen Center for Physics for their hospitality and support where part of this work was carried out.


\begin{thebibliography}{99}
\bibitem{knobel}R. Knobel and A. N. Cleland, Nature {\bf 424}, 291 (2003). 
\bibitem{lahaye}M. D. LaHaye, O. Buu, B. Camarota, and K. C. Schwab, Science {\bf 304}, 74 (2004).
\bibitem{naik}A. Naik, O. Buu, M. D. Lahaye, A. D. Armour, A. A. Clerk, M. P. Blencowe, and K. C. Schwab, Nature {\bf 443}, 193 (2006).
\bibitem{blencoweapl00}M. P. Blencowe and M. N. Wybourne, Appl. Phys. Lett. {\bf 77}, 3845 (2000).
\bibitem{zhangjap02}Y. Zhang and M. P. Blencowe, J. Appl. Phys. {\bf 91}, 4249 (2002).
\bibitem{blencowepr04}M. Blencowe, Phys. Rep. {\bf 395}, 159 (2004).
\bibitem{mozyrskyprl04}D. Mozyrsky, I. Martin, and M. B. Hastings, Phys. Rev. Lett. {\bf 92}, 018303 (2004).
\bibitem{clerknjp05}A. A. Clerk and S. Bennett, N. J. Phys. {\bf 7}, 238 (2005).
\bibitem{blencowenjp05}M. P. Blencowe, J. Imbers, and A. D. Armour, N. J. Phys. {\bf 7}, 236 (2005).  
\bibitem{caves}C. M. Caves, Phys. Rev. D {\bf 26}, 1817 (1982).
\bibitem{braginsky92}V. B. Braginsky and F. Ya. Khalili, {\em Quantum Measurement}, (Cambridge University Press, Cambridge, UK, 1992).
\bibitem{clerkprb04}A. A. Clerk, Phys. Rev. B {\bf 70}, 245306 (2004).
\bibitem{sembanjp07}F. Xue, Y. D. Wang, C. P. Sun, H. Okamoto, H. Yamaguchi, and K. Semba, N. J. Phys. {\bf 9}, 35 (2007).
\bibitem{buksprb06}E. Buks and M. P. Blencowe, Phys. Rev. B {\bf 74}, 174504 (2006).
\bibitem{mizelprl06}X. Zhou and A. Mizel, Phys. Rev. Lett. {\bf 97}, 267201 (2006).
\bibitem{buksprpt06}E. Buks, E. Arbel-Segev, S. Zaitsev, B. Abdo, and M. P. Blencowe, $\mathtt{quant}$-$\mathtt{ph/0610158}$ (unpublished).
\bibitem{kochprl80}R. H. Koch, D. J. van Harlingen, and J. Clarke, Phys. Rev. Lett. {\bf 45}, 2132 (1980).  
\bibitem{kochapl81}R. H. Koch, D. J. van Harlingen, and J. Clarke, Appl. Phys. Lett. {\bf 38}, 380 (1981).
\bibitem{nation??}P. Nation, M. P. Blencowe, and E. Buks, in progress.
\bibitem{truitt07}P. A. Truitt, J. B.  Hertzberg, C. C. Huang, K. L. Ekinci, and K. C. Schwab, Nano Lett., {\bf 7}, 120 (2007).
\bibitem{wineland06} D. J. Wineland, J. Britton, R. J. Epstein, D. Leibfried, R. B. Blakestad, K. Brown, J. D. Jost, C. Langer, R. Ozeri, S. Seidelin, and J. Wesenberg, $\mathtt{quant}$-$\mathtt{ph/0606180}$ (unpublished). 
\bibitem{pacepra93}A. F. Pace, M. J. Collett, and D. F. Walls, Phys. Rev. A {\bf 47}, 3173 (1993).
\bibitem{mancinipra94}S. Mancini and P. Tombesi, Phys. Rev. A {\bf 49}, 4055 (1994).
\bibitem{milburnpra94}G. J. Milburn, K. Jacobs, and D. F. Walls, Phys. Rev. A {\bf 50}, 5256 (1994).
\bibitem{jacobspra99}K. Jacobs, I. Tittonen, H. M. Wiseman, S. Schiller, Phys. Rev. A {\bf 60}, 538 (1999). 
\bibitem{marquardtprpt07}F. Marquardt, J. P. Chen, A. A. Clerk, and S. M. Girvin, $\mathtt{cond}$-$\mathtt{mat/0701416}$ (unpublished). 
\bibitem{blaispra04}A. Blais, R.-S. Huang, A. Wallraff, S. M. Girvin, and R. J. Schoelkopf, Phys. Rev. A {\bf 69}, 062320 (2004). 
\bibitem{wallraffnature04}A. Wallraff, D. Schuster, A. Blais, L. Frunzo, R.-S. Huang, J. Majer, S. Kumar, S. M. Girvin, and R. J. Schoelkopf, Nature {\bf 431}, 165 (2004).
\bibitem{lupascuprl06}A. Lupa\c{s}cu, E. F. C. Driessen, L. Roschier,  C. J. P. M. Harmans,  and J. E. Mooij, Phys. Rev. Lett. {\bf 96}, 127003 (2006).
\bibitem{johanssonjpcm06}G. Johansson, L. Tornberg, V. S. Shumeiko, and G. Wendin, J. Phys.: Condens. Matter {\bf 18}, S901 (2006).
\bibitem{yurkepra84}B. Yurke and J. S. Denker, Phys. Rev. A {\bf 29}, 1419 (1984).
\bibitem{orlando91}T. P. Orlando and K. A. Delin, Chapter 8 of {\em Foundations of Applied Superconductivity}, (Addison-Wesley Publishing Company, Reading, Massachusetts, 1991).
\bibitem{devoret97}M. H.  Devoret, in {\em Quantum Fluctuations (Les Houches Session LXIII)}, edited by  S. Reynaud, E. Giacobino, and J. Zinn-Justin (Elsevier Science, 1997), pp. 351-386.
\bibitem{gardinerpra85}C. W. Gardiner and M. J. Collett, Phys. Rev. A {\bf 31}, 3761 (1985).
\bibitem{gardiner00}C. W. Gardiner and P. Zoller, {\em Quantum Noise: 2nd Edition}, (Springer-Verlag, Berlin, 2000).
\bibitem{yurkejlt07}B. Yurke and E. Buks, J. Lightwave Tech. {\bf 24}, 5054 (2006).
\bibitem{mozyrskyprl02}D. Mozyrsky and I. Martin, Phys. Rev. Lett. {\bf 89}, 018301 (2002).
\bibitem{armourprb04}A. D. Armour, M. P. Blencowe, and Y.  Zhang, Phys. Rev. B {\bf 69}, 125313 (2004).
\bibitem{blencowecontempphys05}M. P. Blencowe, Contemp. Phys. {\bf 46}, 249 (2005).
\bibitem{braginskypla02}V. B. Braginsky and S. P. Vyatchanin, Phys. Lett. A {\bf 293}, 228 (2002).
\bibitem{martinprb04}I. Martin, A. Shnirman, L. Tian, and P. Zoller, Phys. Rev. B {\bf 69}, 125339 (2004).
\bibitem{wilsonprl04}I. Wilson-Rae, P. Zoller, and A. Imamoglu, Phys. Rev. Lett. {\bf 92}, 075507 (2004).
\bibitem{hohbergernature04}C. H\"{o}hberger-Metzger and K. Karrai, Nature {\bf 432}, 1002 (2004).
\bibitem{harrisrsi07}J. G. E. Harris, B. M. Zwickl, and A. M. Jayich, Rev. Sci. Instrum. {\bf 78}, 013107 (2007).
\bibitem{gigannature06}S. Gigan, H. R.  Bohm, M. Paternostro, F. Blaser, G. Langer, J. B. Hertzberg, K. C. Schwab, D. Bauerle, M.  Aspelmeyer, and A.  Zeilinger, Nature {\bf 444}, 67 (2006).
\bibitem{arcizetnature06}O. Arcizet, P.  F. Cohadon, T. Briant, M. Pinard, and A. Heidmann, Nature {\bf 444}, 71 (2006).
\bibitem{vahalaprl06}A. Schliesser, P. DelÕHaye, N. Nooshi, K. J. Vahala, and T. J. Kippenberg, Phys. Rev. Lett. {\bf 97}, 243905 (2006).
\bibitem{corbittprpt06}T. Corbitt, Y. Chen, E. Innerhofer, H. Muller-Ebhardt, D. Ottaway, H. Rehbein, D. Sigg, S. Whitcomb, C. Wipf, and N. Mavalvala, $\mathtt{quant}$-$\mathtt{ph/0612188}$ (unpublished).
\end{thebibliography}
\end{document}